\documentclass[a4paper,11pt]{article}
\pdfoutput=1 

\usepackage{jheppub} 

\usepackage[english]{babel}
\usepackage{csquotes}
\usepackage[T1]{fontenc}
\usepackage{amsmath,bm}
\usepackage{amsfonts}
\usepackage{algorithm}
\usepackage{algpseudocode}
\usepackage{bbm}

\usepackage{amssymb}
\usepackage{subfigure}
\usepackage{mathtools}
\usepackage{listings}

\usepackage{blindtext}

\usepackage{hyperref}
\usepackage{braket}
\usepackage{blindtext}
\usepackage{graphicx}
\usepackage{fancyhdr}
\usepackage{feynmp-auto}
\usepackage{tikz}
\usepackage{marvosym}
\usepackage{jheppub}
\usepackage{booktabs} 	
\usetikzlibrary{arrows, shapes.geometric, arrows.meta, shapes, decorations.pathreplacing, fit, patterns, patterns.meta}

\definecolor{Rcolor}{HTML}{E99595}
\definecolor{Gcolor}{HTML}{C5E0B4}
\definecolor{Bcolor}{HTML}{9DC3E6}
\definecolor{Ycolor}{HTML}{FFE699}

\tikzstyle{expr} = [circle, minimum width=1.8cm, minimum height=2.8cm, text centered, align=center, inner sep=0, draw,font=\LARGE]
\tikzstyle{txt_huge} = [align=center, font=\Huge, scale=1]
\tikzstyle{txt} = [align=center, font=\Huge]
\tikzstyle{cinn} = [double arrow, double arrow head extend=0cm, double arrow tip angle=130, shape border rotate=90, inner sep=0, align=center, minimum width=3.1cm, minimum height=3.3cm, fill=Bcolor, draw,font=\Huge]
\tikzstyle{cinn_black} = [cinn, minimum height=3.5cm, fill=black]
\tikzstyle{arrow} = [thick,-{Latex[scale=1.0]}, line width=0.2mm, color=black]
\tikzstyle{line} = [thick, line width=0.2mm, color=black]
\tikzstyle{loss} = [rectangle, align=center,  minimum width=1.8cm, minimum height=1.5cm,fill=Rcolor,font=\LARGE, rounded corners]
\tikzstyle{xt} = [rectangle, align=center,  minimum width=3cm, minimum height=1.5cm,fill=Gcolor,font=\Large, rounded corners]
\tikzstyle{xts} = [rectangle, align=center,  minimum width=1cm, minimum height=1.5cm,fill=Gcolor,font=\Large, rounded corners]

\newcommand\one{\leavevmode\hbox{\small1\normalsize\kern-.33em1}}

\newcommand{\FastJet}{\textsc{FastJet}\xspace}
\newcommand{\Pythia}{\textsc{Pythia}\xspace}
\newcommand{\Delphes}{\textsc{Delphes}\xspace}
\newcommand{\Herwig}{\textsc{Herwig}\xspace}

\newcommand{\arXiv}[2][]{%
	\ifthenelse{\equal{#1}{}}%
	{\href{http://arxiv.org/abs/#2}{arXiv:#2}}%
	{\href{http://arxiv.org/abs/#2}{arXiv:#2~[#1]}}}

\def\slashchar#1{\setbox0=\hbox{$#1$}           
   \dimen0=\wd0                                 
   \setbox1=\hbox{/} \dimen1=\wd1               
   \ifdim\dimen0>\dimen1                        
      \rlap{\hbox to \dimen0{\hfil/\hfil}}      
      #1                                        
   \else                                        
      \rlap{\hbox to \dimen1{\hfil$#1$\hfil}}   
      /                                         
   \fi}

\newcommand{\tikznode}[2]{%
\ifmmode%
\tikz[remember picture,baseline=(#1.base),inner sep=0pt] \node (#1) {$#2$};%
\else
\tikz[remember picture,baseline=(#1.base),inner sep=0pt] \node (#1) {#2};%
\fi}

\def\mathswitchr#1{\relax\ifmmode{\mathrm{#1}}\else$\mathrm{#1}$\xspace\fi}
\def\mathswitch#1{\relax\ifmmode#1\else$#1$\xspace\fi}

\newcommand{\pdsbrr}{\mathswitch{p_{d}(x)}}
\newcommand{\pdsrr}{\mathswitch{p_{d,s}(x)}}
\newcommand{\pdsrrg}{\mathswitch{p_{d,s}(x)_{g}}}

\newcommand{\pdsgrg}{\mathswitch{p_{d,s}(y)_{r}}}
\newcommand{\pdsgg}{\mathswitch{p_{d,s}(y)}}

\newcommand{\pmcsrrg}{\mathswitch{p_{\text{MC}, s}(x)_{g}}}
\newcommand{\pmcbrr}{\mathswitch{p_{\text{MC},b}(x)}}

\newcommand{\pmcsgrg}{\mathswitch{p_{\text{MC},s}(y)_{r}}}

\newcommand{\pdsg}{\mathswitch{p_{d,s}(y)}}
\newcommand{\pdsr}{\mathswitch{p_{d,s}(x)}}
\newcommand{\pmcsg}{\mathswitch{p_{\text{MC},s}(y)}}
\newcommand{\pmcsr}{\mathswitch{p_{\text{MC},s}(x)}}
\newcommand{\pmcbr}{\mathswitch{p_{\text{MC},b}(x)}}
\newcommand{\pmcsrrgbar}{\mathswitch{p_{\text{MC},s}(x)_{\bar{g}}}}
\newcommand{\pdsrrgbar}{\mathswitch{p_{d,s}(x)_{\bar{g}}}}
\newcommand{\pmcsgrbarg}{\mathswitch{p_{\text{MC},s}(y)_{\bar{r}}}}
\newcommand{\pmcsrgrg}{\mathswitch{p_{\text{MC},s}(x,y)_{r,g}}}

\newcommand{\punfold}{\mathswitch{p_{d, \text{unfold}}(y)_{r}}}
\newcommand{\pnunfold}{\mathswitch{p^n_{d, \text{unfold}}(y)_{r}}}
\newcommand{\pmcsgrr}{\mathswitch{p(x|y)}}
\newcommand{\pmcsrg}{\mathswitch{p_{\text{MC},s}(y|x)}}

\newcommand{\pdsrg}{\mathswitch{p_{d,s}(y|x)}}

\title{\boldmath Analysis-ready Generative Unfolding}

\author[a,b]{Anja Butter,}
\author[a]{Nathan Huetsch,}
\author[c]{Vinicius Mikuni,}
\author[d,e]{Benjamin Nachman,}
\author[a]{and Sofia Palacios Schweitzer}

\affiliation[a]{Institut für Theoretische Physik, Universität Heidelberg, Germany}
\affiliation[b]{LPNHE, Sorbonne Universit\'e, Universit\'e Paris Cit\'e, CNRS/IN2P3, Paris, France}
\affiliation[c]{National Energy Research Scientific Computing Center, Berkeley Lab, Berkeley 94720, USA}
\affiliation[d]{Fundamental Physics Directorate, SLAC National Accelerator Laboratory,
Menlo Park, CA 94025, USA}
\affiliation[e]{Department of Particle Physics and Astrophysics, Stanford University,
Stanford, CA 94305, USA}

\abstract{Machine Learning (ML)-based unfolding methods have enabled high-dimensional and unbinned differential cross section measurements. While a suite of such methods has been proposed, most focus exclusively on the challenge of statistically removing resolution effects.  In practice, unfolding methods must also account for impurities and finite acceptance and efficiency effects. In this paper, we extend a class of unfolding methods based on generative ML to include the full suite of effects relevant for cross section measurements. Our new methods include fully generative solutions as well as generative-discriminative hybrid approaches (GenFoldG and GenFoldC). We demonstrate these new techniques in both Gaussian and simulated LHC examples.  Overall, we find that both methods are able to accommodate all effects, thus adding a complementary and analysis-ready method to the unfolding toolkit.}

\begin{document}
\maketitle
\flushbottom
\section{Introduction}
\label{sec:intro}

Particle physics experiments seek to understand the properties of and the interactions between fundamental particles.  To do so, complex detectors measure outgoing particle properties and the underlying phenomena are inferred.  The inference step is either performed at the detector-level by forward-folding predictions with simulations of the detector or by first statistically removing the effects of the detector -- also known as unfolding -- before comparing the data to particle-level predictions.  In either case, the classical approach is to compress and discretize the high-dimensional data into binned, differential cross sections represented as histograms.  This has enabled nearly all measurements to date, but significantly reduces the statistical power of the data.

Machine Learning (ML) has provided a solution to this challenge by enabling high-dimensional and unbinned inference~\cite{Cranmer:2019eaq}.  For unfolding~\cite{Arratia:2021otl, Huetsch:2024quz,Canelli:2025ybb}, this means that binnings and even observables can be chosen after the measurement, ushering in a new analysis that increases the breadth and longevity of particle physics science results. One class of unbinned unfolding methods based entirely on discriminative (classifier based) ML called OmniFold~\cite{Andreassen:2019cjw,Andreassen:2021zzk} has already been employed for a diverse set of differential cross section measurements and studies across experiments including ATLAS~\cite{ATLAS:2024xxl,ATLAS:2025qtv}, CMS~\cite{Komiske:2022vxg,CMS:2025sws}, LHCb~\cite{LHCb:2022rky}, H1~\cite{H1:2021wkz,H1:2023fzk,H1:2024mox,h1_full_phase_space}, STAR~\cite{Song:2023sxb,Pani:2024mgy}, T2K~\cite{Huang:2025ziq}, and ALEPH~\cite{Badea:2025wzd}. 

Given the ill-posed nature of unfolding, it is essential to have multiple methods.  A number of generative ML-based unfolding techniques have also been proposed to enable unbinned, high-dimensional cross section measurements.  Such approaches include methods that generate particle-level events based on detector-level events using conditional generative networks~\cite{Datta:2018mwd,Bellagente:2019uyp,Bellagente:2020piv,Howard:2021pos,Backes:2022sph,Ackerschott:2023nax,Shmakov:2023kjj,Shmakov:2024gkd,Pazos:2024nfe,Favaro:2025psi} 
or distribution mapping~\cite{Diefenbacher:2023wec,Butter:2023ira,
Butter:2024vbx}, as well as approaches using surrogates of the detector simulation to adjust the particle-level distribution such that the forward-folded prediction best reproduces the observed data~\cite{Alanazi:2020jod,Vandegar:2020yvw,Butter:2025via}. The former class of techniques that generate particle-level events with detector-level inputs are inherently dependent on the starting particle-level simulation used to train the generative model.  This dependence can be mitigated by iterating the training of the generative model~\cite{Backes:2022sph}, removing one of the largest sources of measurement bias.

Generative unfolding proposals have so far focused exclusively on the statistical removal of resolution effects\footnote{The full protocol for classifier-based approaches is discussed in Ref.~\cite{Andreassen:2021zzk,Falcao:2025jom} and in the experimental papers cited above.}. While critical, this is not the only effect that is relevant in practice. Additional tasks include the removal of background processes, correcting for events that are registered at detector-level, but not particle-level (fakes\footnote{Fakes and background are conceptually similar.  Indeed, one solution to addressing backgrounds is to treat them as fakes. We treat them separately here because they are usually estimated and processed differently in practice.}), and correcting for events that are registered at particle-level, but not detector-level (misses).  Each of these must also be processed unbinned and in many dimensions for the final result to retain these properties.

In this paper, we extend iterative, generative unfolding methods to include sample impurity (backgrounds), acceptance (fakes), and efficiency (misses) effects. In addition to building on the iterative unfolding algorithm introduced in Ref.~\cite{Backes:2022sph}, we also develop a new iterative approach. We start by introducing the methodology of our generative unfolding pipeline in Sec.~\ref{sec:pipeline} and formulate correction algorithms which can be implemented either with classifiers or generative neural networks. We present each classifier- and generator-based correction thoroughly with the necessary mathematical and computational formalism as well as illustratively, with a Gaussian example in Sec.~\ref{sec:gaussian_toy}.
We then proceed to unfold a 6-dimensional simulated dataset, building on a well-studied benchmark that now includes background. This dataset consists of single jet substructure observables and is unfolded in Sec.~\ref{sec:omnifold_data} utilizing the full generative unfolding pipeline for the first time on a physics example. 
%

\section{Methodology}
\label{sec:pipeline}


Emulated events in particle physics are created by chaining together a suite of Monte Carlo simulation tools.  The set of tools consists of multiple parts: hard scattering, parton showering, hadronization, and the simulation of the detector response. Depending on the physics analysis we can unfold for instance to parton level, i.e. the final state of the hard scattering, or particle level, i.e. the representation of an event before it enters the detector simulation. The data themselves are given at detector-level or reco-level, which corresponds to the final stage of the simulation, when event reconstruction is applied to the output of the detector simulation.  While unfolding of hadronization and/or parton shower effects has been performed in previous studies~\cite{Bellagente:2020piv, Bellagente:2021yyh, Shmakov:2024gkd}, we will focus in this paper on unfolding detector effects only.
%

\begin{figure}[t]
    \centering
    \begin{tikzpicture}[node distance=2cm, scale=0.7, every node/.style={transform shape}]

\fill[red, opacity=0.1] (-1,0) circle (2);
\fill[blue, opacity=0.1] (1,0) circle (2);

\draw[red, opacity=0.5] (-1,0) circle (2);
\draw[blue, opacity=0.5] (1,0) circle (2);

\node at (-2.2, 0) {\Large $p(x)_{r,\bar{g}}$};
\node at (2.2, 0) {\Large $p(x)_{\bar{r},g}$};
\node at (0, 0) {\Large $p(x)_{r,g}$};
\draw[thick, rounded corners] (-4.5,-2.3) rectangle (4,3.9);
\node[below right] at (4,-2.5) {\Large $x$};

\begin{scope}[shift={(-4,3)}]
  \draw[fill=red, opacity=0.3, rounded corners] (0,0) rectangle (0.4,0.4);
  \node[right=2pt] at (0.5,0.2) {\Large reco-level selection};

  \draw[fill=blue, opacity=0.3, rounded corners] (0,-0.6) rectangle (0.4,-0.2);
  \node[right=2pt] at (0.5,-0.4) {\Large gen-level selection};
\end{scope}
\end{tikzpicture}
    \caption{Graphical description of different reco-level phase spaces.}
    \label{fig:venn_diagram}
\end{figure}
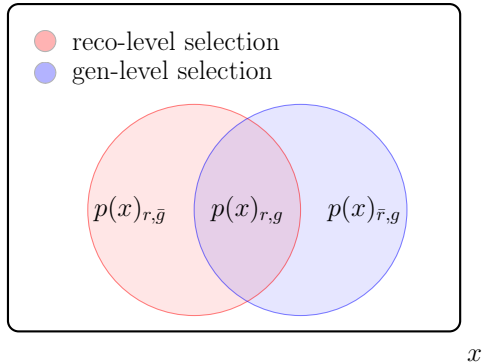

\begin{figure}[t]
    \begin{tikzpicture}[node distance=2cm, scale=0.34, every node/.style={transform shape}]

\node (pdata) [txt] {$\pdsbrr$};
\node (background_b) [cinn_black, right of = pdata, xshift=3cm] {};
\node (background) [cinn, right of = pdata, xshift=3cm] {$- b$};
\node (psignal) [txt, right of = background, xshift=3cm] {$\pdsrr$};
\node (scissors_b) [cinn_black, right of = psignal, xshift=3cm]{};
\node (scissors) [cinn, right of = psignal,fill=Gcolor, xshift=3cm]{$\times \delta$};
\node (pacc) [txt, right of = scissors, xshift=3cm]{$\pdsrrg$};
\node (unfolder_b) [cinn_black, right of = pacc, xshift=3cm] {};
\node (unfolder) [cinn, right of = pacc, xshift=3cm, fill=Rcolor] {Unfold};
\node (punfold) [txt, right of = unfolder, xshift=3cm]{$\pdsgrg$};
\node (eff_b) [cinn_black, right of = punfold, xshift=3cm] {};
\node (eff) [cinn, right of = punfold, xshift=3cm, fill= Ycolor] {$\times {\epsilon}$};
\node (peff) [txt, right of = eff, xshift=3cm]{$\pdsgg$};


\node (background_subtractor) [txt, above of = background_b, yshift=1cm]{1. Step: \\ Background Subtraction};
\node (unfolding) [txt, above of = unfolder_b, yshift=1cm]{3. Step: \\ Unfolding};
\node (iterating) [txt, below of = unfolder_b, yshift=-3.5cm]{4. Step: \\ Prior Removal};
\node (efficieny) [txt, above of = eff_b, yshift=1cm]{5. Step: \\ Efficiency correction};
\node (acceptance) [txt, above of = scissors, yshift=1cm]{2. Step: \\ Acceptance correction};

\draw [arrow, color=black] (pdata.east) -- (background.west);
\draw [arrow, color=black] (background.east) -- (psignal.west);
\draw [arrow, color=black] (psignal.east) -- (scissors.west);
\draw [arrow, color=black] (unfolder.east) -- (punfold.west);
\draw [arrow, color=black] (punfold.east) -- (eff.west);
\draw [arrow, color=black] (eff.east) -- (peff.west);
\draw [arrow, color=black] (pacc.east) -- (unfolder.west);
\draw [arrow, color=black] (scissors.east) -- (pacc.west);

\draw [arrow, dashed, color=black] (punfold) to [out=-120, in=-60] (pacc);

\end{tikzpicture}
    \caption{Visualization of the full generative unfolding pipeline to infer $\pdsgg$ all the way from $\pdsbrr$ in a five step approach.}
    \label{fig:unfolding_pipeline}
\end{figure}
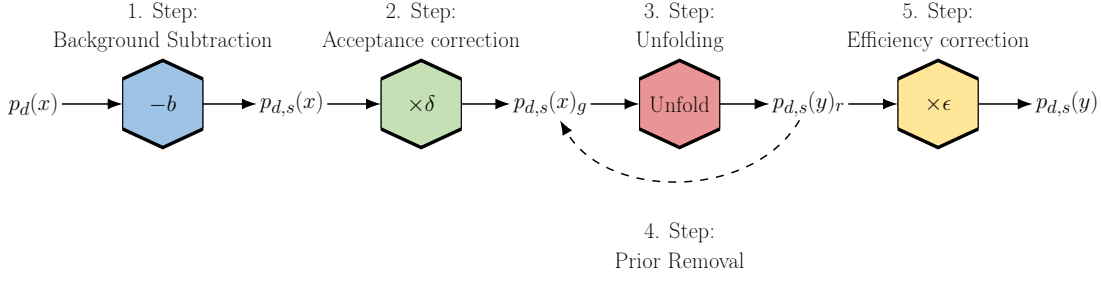

Since we simulate the entire chain, pairs of events at both gen- and reco-level are available for further analysis. Using these pairs, we can learn to simulate both the detector effects (from gen-level to reco-level) as well as to estimate the unfolding response (from reco-level to gen-level).
Starting from the signal simulation of gen-level collisions $y_{s}\sim\pmcsg$, we determine reco-level objects $x_{s}\sim\pmcsr$ from the detector simulation. In addition to the signal process, background processes can also be present in the data, resulting in samples $x_{b} \sim \pmcbr$. Simulations of the background process at gen-level are not needed since we subtract the background from the data before unfolding\footnote{This is standard, in part because it simplifies the next steps. However, it usually requires a strict event selection and this results in large acceptance effects.  Unbinned unfolding makes it possible to apply nontrivial event selections after unfolding so this order may be revisited in the future.}. Acceptance and efficiency effects arise from events whose pairs are not present either at reco- or gen-level. This is caused by the choice of fiducial region under study and the finite resolution of the detector response.  In order to model acceptance and efficiency effects, we keep track of a flag that identifies whether a given event has passed the reco-level, gen-level, or both  selections. For example, reco-level events that passed all selection criteria are sampled from $x \sim p(x)_{r,g}$ as schematically visualized in Fig.~\ref{fig:venn_diagram}. On reco- (gen-)level, we only consider events that passed reco- (gen-)level selection and therefore omit the dependence on $r \; (g)$. 
After applying the reco-level selection, both signal and background processes are present in the data. These background events need to be statistically removed in order to determine the unfolded signal distribution while events not passing the fiducial selection need to be accounted. At the end, the unfolding algorithm should reproduce the gen-level signal distribution defined within a particular fiducial region $\pdsgg $.
\clearpage
\noindent The full generative unfolding pipeline, as illustrated in Fig.~\ref{fig:unfolding_pipeline}, consists of
\begin{enumerate}
    \item statistically subtracting background events from observed data;
    \item accounting for acceptance effects, i.e.~correcting for
    measured events that are not present in the final fiducial region;
    \item estimating the gen-level distribution from the reco-level data (previously, the focus of generative unfolding studies);
    \item iteratively removing the dependence on the prior simulation;
    \item accounting for efficiency effects, i.e.~accounting for events present in the final fiducial region but not in the reconstructed region.
\end{enumerate}

\subsection{Background Subtraction}
\label{sec:background_sub}
\subsubsection*{Classifier-based}
\begin{algorithm}
\caption{Background-subtraction classifier}
\label{alg:background_classifier}
\begin{algorithmic}[1]
\Require Reco-level data $x_{d}\sim \pdsr$ with weights $1$ and reco-level background $x_{\text{MC}, b} \sim \pmcbr$ with weights $-1$
\State Construct training dataset as follows:
\Statex \hspace{\algorithmicindent} $\bullet$ Assign labels 1 to $\{x_{d}, x_{\text{MC}, b}\}$
\Statex \hspace{\algorithmicindent} $\bullet$ Assign labels 0 to $\{x_{d}\}$
\State Train classifier to distinguish samples of different labels with a BCE loss
\end{algorithmic}
\end{algorithm}
We follow the classifier-based event subtraction in Alg.~\ref{alg:background_classifier}, which was initially proposed in Ref.~\cite{Andreassen:2021zzk} based on Ref.~\cite{Nachman:2020fff}. 
Once the classifier converges, the likelihood ratio 
\begin{align}
    \nu(x) \equiv \frac{\pdsbrr - \pmcbrr}{\pdsbrr} = \frac{C(x)}{1-C(x)}\,,
    \label{eq:backgrond_weights}
\end{align}
can be related to classifier output $C(x)$. We train with a balanced classifier as we know the total number of background events, i.e.~we can normalize the sum of weights to the expected number of signal events after training. By weighing each event $x_{d}$ with the corresponding weight $\nu(x)$ we probabilistically subtract the background from the observed data. 
Similarly to bin-wise background subtraction, we need to assume that we have access to reliable estimates of the background process. This can either be accomplished with reliable simulations or using data-driven estimates of the background.
An unfolded distribution can then be background corrected by the learned weight $\nu(x) \rightarrow \nu(y)$ which can be pulled to the corresponding gen-level event.
\subsubsection*{Generator-based}
\begin{algorithm}
\caption{Background-subtraction generator}
\label{alg:background_generator}
\begin{algorithmic}[1]
\Require Reco-level data $x_{d} \sim \pdsr$ and reco-level background $x_{\text{MC}, b} \sim \pmcbr$
\State Construct training dataset as follows:
\Statex \hspace{\algorithmicindent} $\bullet$ Assign weights $1$ to data samples $x_{d}$
\Statex \hspace{\algorithmicindent} $\bullet$ Assign weights $-1$ to background samples samples $x_{\text{MC}, b}$

\State Train a generative model to generate data using the combined sample as the training dataset
\end{algorithmic}
\end{algorithm}

Background subtraction using generative models has been studied in the past with the use of GANs~\cite{Butter:2019eyo2} or normalizing flows~\cite{Das:2023bcj,Cheng:2025ewj}. The simplest method is to follow a similar approach to the classifier strategy, where the training sample can be constructed based on the available data combined with simulated background samples with negative weights. In Alg.~\ref{alg:background_generator}, the strategy to learn the signal distribution using a generative model is described. Using the negative weights during the training leads to a model capable of generating the background-subtracted data at inference time.  
The strategy decouples from the type of generative model used. The density approximated by the generative model is then

\begin{equation}
    p_{\text{model},s}(x) = \pdsbrr - \pmcbrr.
    \label{eq:bkg_sub_gen}    
\end{equation}
\subsection{Acceptance Correction}
\label{sec:acceptance}
\begin{algorithm}
\caption{Acceptance classifier}
\label{alg:acceptance_classifier}
\begin{algorithmic}[1]
\Require Reco-level signal MC from $x_{g} \sim \pmcsrrg$ and from $x_{\bar{g}} \sim \pmcsrrgbar$
\State Construct training dataset as follows:
\begin{itemize}
    \item Assign labels 1 to $\{x_{g}\}$ 
    \item Assign labels 0 to $\{x_{\bar{g}}\}$
\end{itemize}
\State Train classifier to distinguish samples of different labels with a BCE loss
\end{algorithmic}
\end{algorithm}
Acceptance effects arise from a mismatch between the data measured and the fiducial region presented after unfolding. In particular, events that are reconstructed in the experiment can lie outside of the fiducial region of interest. Since the detector response is stochastic, we need to statistically learn the rate a reco-level event is also present at gen-level given a fiducial region definition and given the observed reco-level properties.


The goal is to learn the conditional acceptance probability given a reco-level events $x$, which we can rewrite using Bayes' theorem to
\begin{align}
    \label{eq:acceptance_prob}
    p_{d,s}(g|x) = \frac{p(g)\pdsrrg}{p(g)\pdsrrg + p(\bar{g})\pdsrrgbar } \;, 
\end{align}
where $p(g)$ is the unconditional acceptance probability. It is estimated by the number of events in simulation that passed gen-level selection $N_g$ divided by the total number of events
\begin{align}
    p(g) = \frac{N_g}{N_\text{total}} \;.
\end{align}
However, we neither have access to $\pdsrrg $ nor to samples from that distribution. Therefore, we estimate the ratio in Eq.~\eqref{eq:acceptance_prob} by using the simulation, assuming the same acceptance probability $p(g)$ in data and MC. This can be done by training a classifier according to Alg.~\ref{alg:acceptance_classifier}. After convergence, the ratio can then be approximated by the classifier output 
\begin{align}
    \delta(x) \equiv  \frac{p(g)\pmcsrrg}{p(g)\pmcsrrg + p(\bar{g})\pmcsrrgbar } \; .
    \label{eq:acceptance}
\end{align}
Once unfolded, we can pull the acceptance correction to gen-level $\delta(x) \rightarrow \delta(y)$ or use the same classifier to determine which events at reconstruction level are likely to be found in the fiducial region of interest.
\subsection{Efficiency Correction}
\label{sec:efficiency}
\begin{algorithm}
\caption{Efficiency classifier}
\label{alg:efficiency_classifier}
\begin{algorithmic}[1]
\Require Gen-level signal MC from $y_{r} \sim \pmcsgrg$ and from $y_{\bar{r}} \sim \pmcsgrbarg$
\State Construct training dataset as follows:
\begin{itemize}
    \item Assign labels 1 to $\{y_{r}\}$ 
    \item Assign labels 0 to $\{y_{\bar{r}}\}$ 
\end{itemize}
\State Train classifier to distinguish samples of different labels with a BCE loss
\end{algorithmic}
\end{algorithm}
Conversely, we also need to account for the events that belong in the reported fiducial region but fail the reconstruction. Similarly to the acceptance classifier, we can train an efficiency classifier that instead takes gen-level inputs to estimate the rate gen-level events are reconstructed.

In analogy to the acceptance classifier, we train a classifier following 
Alg.~\ref{alg:efficiency_classifier}. This was initially proposed in Ref.~\cite{DiBello:2020ppq,Heimel:2023mvw}. Once converged, the reciprocal classifier output $\epsilon(y)$ can be translated to 
\begin{align}
    \epsilon(y) \equiv  \frac{p(\bar{r})\pmcsgrbarg + p(r)\pmcsgrg}{p(r)\pmcsgrg} \; ,
    \label{eq:efficiency}
\end{align}
where we estimate the efficiency probability in simulation by the ratio of events passing the reco-level selection and the total number of events
\begin{align}
    p(r) = \frac{N_r}{N_\text{total}} \;.
\end{align}
Depending on the final unfolding strategy, the application of the acceptance and efficiency classifier change. However, all methods present in this work use classifiers to estimate the acceptance and efficiency corrections.

\subsection{Generative Unfolding}
\label{sec:gen_unfolding}
A generative neural network is trained to learn to produce samples from the posterior distribution $p_{\text{MC},s}(y|x_{d,s})$. During training, paired simulated samples ${x,y} \sim \pmcsrgrg$ passing both reco- and gen-level selection are used. 
After training we can compute $\punfold$ by 
\begin{align}
    \punfold = \int dx \; p_{\text{MC},s}(y|x_{d,s}) p_{d,s}(x)_g \;.
\end{align}

%
\subsection{Removing Prior Simulation Dependence}
\label{sec:prior}
The learned posterior is by construction dependent on the initial simulation used for the training, namely the gen-level distribution realized in the training data $\pmcsg$. We can reduce the prior dependence by iteratively using the obtained unfolded distribution as the new prior. Technically this can be realized with two distinct methods.
\subsubsection*{Classifier-based}
\begin{algorithm}
\caption{Iterative generative unfolding with classifier update}
\label{alg:iterative_classifier}
\begin{algorithmic}[1]
\Require data after background subtraction $x_d$, and acceptance and efficiency classifiers $\delta(x)$ and $\epsilon(y)$

\State Train unfolding model $\pmcsrg$ with paired signal MC $(x,y) \sim \pmcsrgrg$
\State Unfold data and correct for acceptance $y_d \sim \delta (y) \punfold$

\State Construct training dataset as follows:
\begin{itemize}
    \item Assign labels 1 to $\{y_{d}\}$
    \item Assign labels 0 to $\{y_\text{MC}\}$
\end{itemize}
\State Train classifier to distinguish samples of different labels with a BCE loss
\State Evaluate classifier and compute weights $w(y)$
\State Update $\pmcsgrg \rightarrow w(y) \pmcsgrg$ and iterate back to 1.
\end{algorithmic}
\end{algorithm}
To iteratively mitigate the prior dependence using a classifier, we follow Ref.~\cite{Backes:2022sph}. The training procedure is sketched in Alg.~\ref{alg:iterative_classifier}. After training a generative network to learn the conditional density distribution $p_{\text{MC},s}(y|x)$, it uses the measured data $x_d$ with background corrections to iteratively update the unfolded response. A classifier is trained to distinguish between the unfolded distribution $\delta(y)\punfold$ and $\pmcsgrg$. After convergence, the classifier output $E(y)$ can be related to
\begin{align}
    w(y) \equiv  \frac{\nu(y) \delta(y)p_{d,s}(y)}{\pmcsgrg} = \frac{E(y)}{1-E(y)} \;.
    \label{eq:prior_weight}
\end{align}
We then update the generative unfolding algorithm using sampled paired events from the new joint distribution $x,y \sim w(y)  p_{\text{MC},s}(x,y)_{r,g}$. 
This iterative process is repeated $n$ times, after which the unfolded distribution $\pnunfold$ approximates the true distribution $\pdsgrg$. 

Next, we multiply the weights associated with the unfolded distribution by $\epsilon(y)$ to correct for efficiency effects.

We then arrive at the final result of the unfolding algorithm (which we now call GenFoldC), which for the sake of readability, we define as
\begin{align}
   p_{\text{GenFoldC}}(y) \equiv \epsilon(y)\delta(y)\pnunfold \;.
    \label{eq:final_unfolding}
\end{align}
\subsubsection*{Generator-based}
%
\begin{algorithm}
\caption{Iterative generative unfolding with generator update}
\label{alg:iterative_generator}
\begin{algorithmic}[1]
\Require initial models $\pdsg = \pmcsg$, $\pdsrg =\pmcsrg$, $\pmcsgrr$, data after background subtraction $x_d$, and acceptance and efficiency classifiers $\delta(x)$ and $\epsilon(y)$

\State Create $N_{\emptyset} = \frac{N_d\epsilon}{1-\epsilon}$ empty events $x_\emptyset$ and combine with the $N_d$ non-empty data entries $x_d$

\State Generate $N_d + N_{\emptyset}$ events from $y\sim \pdsg$ 
\State Sample from a Bernoulli distribution using $\delta(x_d)$ as probabilities and convert gen-level events into empty entries depending on the sampled value

\State Generate reco-level events $x \sim p(x|[y,y_\emptyset])$
\State Sample from a Bernoulli distribution using $\epsilon(y)$ as probabilities and convert reco-level events into empty entries depending on the sampled value

\State Update $\pdsrg$ using the generated pairs ${x,y}$ for entries where both $x$ and $y$ are not empty
\State Sample $y_d\sim p_{d,s}(y|[x_s,x_\emptyset])$
\State Update $\pdsg$ using events $y_d$ and iterate back to 2.
\end{algorithmic}
\end{algorithm}
Our proposed alternative to the classifier-based method is to use generative models at every step of the iterations as sketched in Alg.~\ref{alg:iterative_generator}. We use the initial simulation to learn the gen-level distribution $\pmcsg$ and the detector response $\pmcsgrr$, similar to Ref.~\cite{Butter:2025via}. The combination of the two allows us to sample from the joint distribution $p_{\text{MC},s}(x,y)$. Since we assume the detector response is universal, each unfolding iteration updates $\pmcsgrg$ while $\pmcsgrr$ remains fixed. After generating samples from the joint distribution, we train the initial unfolding model $\pmcsrg$ using the pairs  $x,y \sim p_{\text{MC},s}(x,y)$. From our unfolding model, we can now generate initial unfolded samples $y_{d,s}\sim p_{\text{MC},s}(y|x_{d,s})$ that are used to update the density $p_{\text{unfold}}(y)$. We repeat this process for $n$ iterations until convergence, with the final set of unfolded events generated directly from the learned $p^n_{\text{unfold}}(y)$ after the last iteration. \\
Since we do not use weights to adjust the probability densities, we need to include the acceptance and efficiency effects as part of the updates. This is achieved by the introduction of empty events during the training to mimic the behavior of events that are lost due to detector effects (acceptance) or in the converse process (efficiency). The implementation of the empty events is done through the addition of events with all entries set to a specific value, such as 0 for example. The total amount of empty events added to the initial data is determined by the estimated efficiency from the simulation. We estimate the number of empty events at gen-level by using the acceptance classifier output evaluated across all reco-level events. From the classifier output, we use the predicted probabilities $\delta(x)$ to sample from a Bernoulli distribution that either accepts or rejects the gen-level pair associated to event $x$. 
The final results of the GenFoldG algorithms are unweighted events that account for acceptance, efficiency, and background effects following 
\begin{align}
    p_\text{GenFoldG}(y) \approx \pdsgg \; .
\end{align}
%
%


\section{Gaussian Example}
\label{sec:gaussian_toy}
We show the interplay of the multiple steps required to unfold data by first considering the Gaussian example from Ref.~\cite{Andreassen:2021zzk}.

We sample gen-level, signal (pseudo-)data, gen-level signal MC and reco-level background MC from Gaussian distributions
\begin{align}
    \pmcsg &= \mathcal{N}(0,1)  \; , \qquad 
    \pdsg  = \mathcal{N}(0.2,0.8) \;, \qquad
    \pmcbr = \mathcal{N}(0,1.2) \; .
    \label{eq:toy_model}
\end{align}
%
To emulate detector effects we smear these events with a Gaussian kernel $\mathcal{N}(0, 0.5)$, such that the corresponding reco-level distribution for MC and data are given by 
\begin{align}
    \pmcsr &= \mathcal{N}(0,\sqrt{1^2 + 0.5^2 })  \; , \qquad 
    \pdsr  =\mathcal{N}(0.2,\sqrt{0.8^2 + 0.5^2 })\; .
    \label{eq:toy_model_reco}
\end{align}
In total, there is 10\% background contamination following Eq.~\eqref{eq:toy_model}. 
Furthermore, we study the case of uniformly distributed efficiency and acceptance effects,~i.e. we randomly drop 10\% of the data and simulation on reco- and gen-level, respectively.

\begin{figure}[t]
    \centering    
    \includegraphics[width=0.45\textwidth]{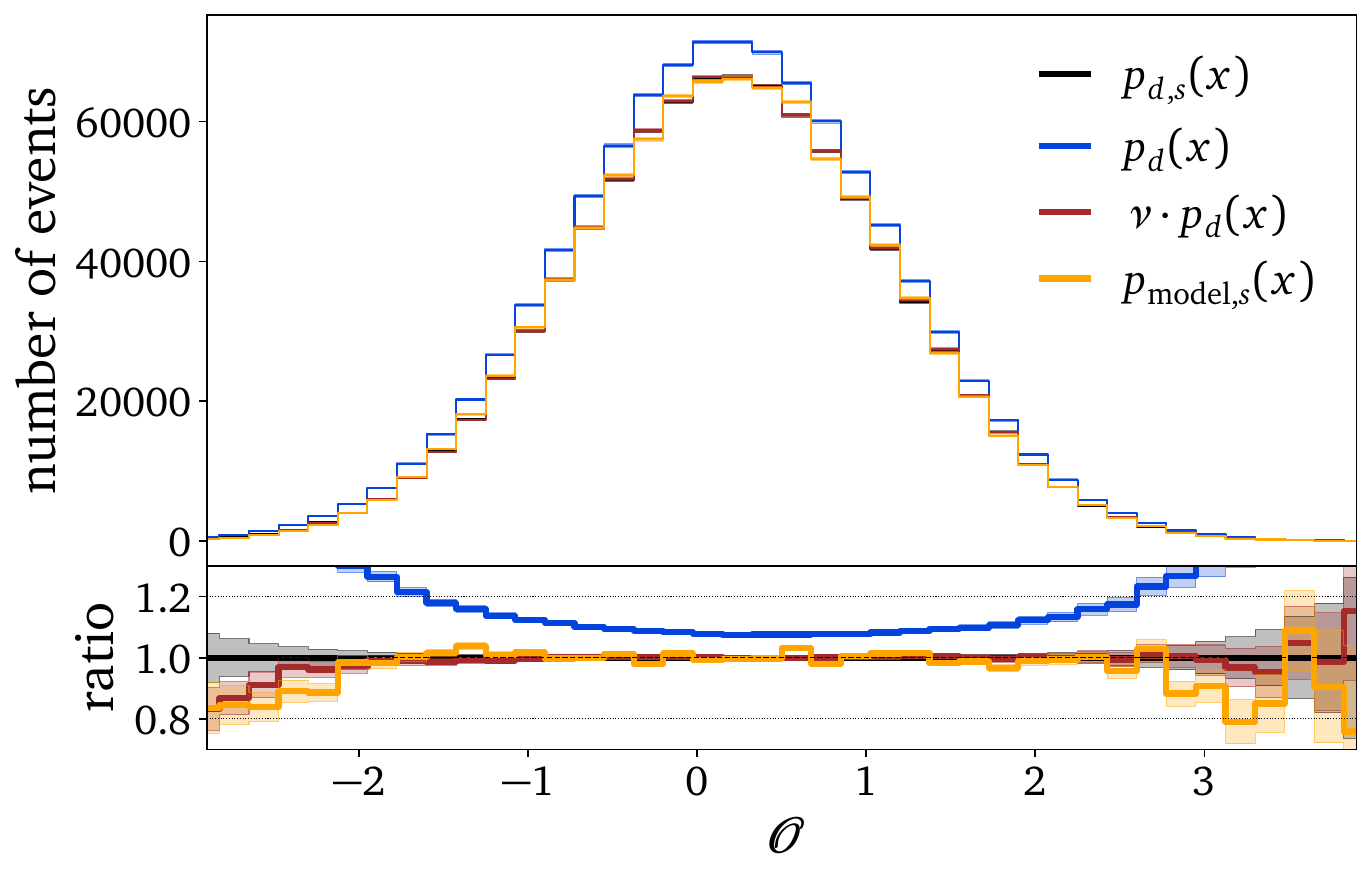}
    \hspace{0.5cm}
    \includegraphics[width=0.45\textwidth]{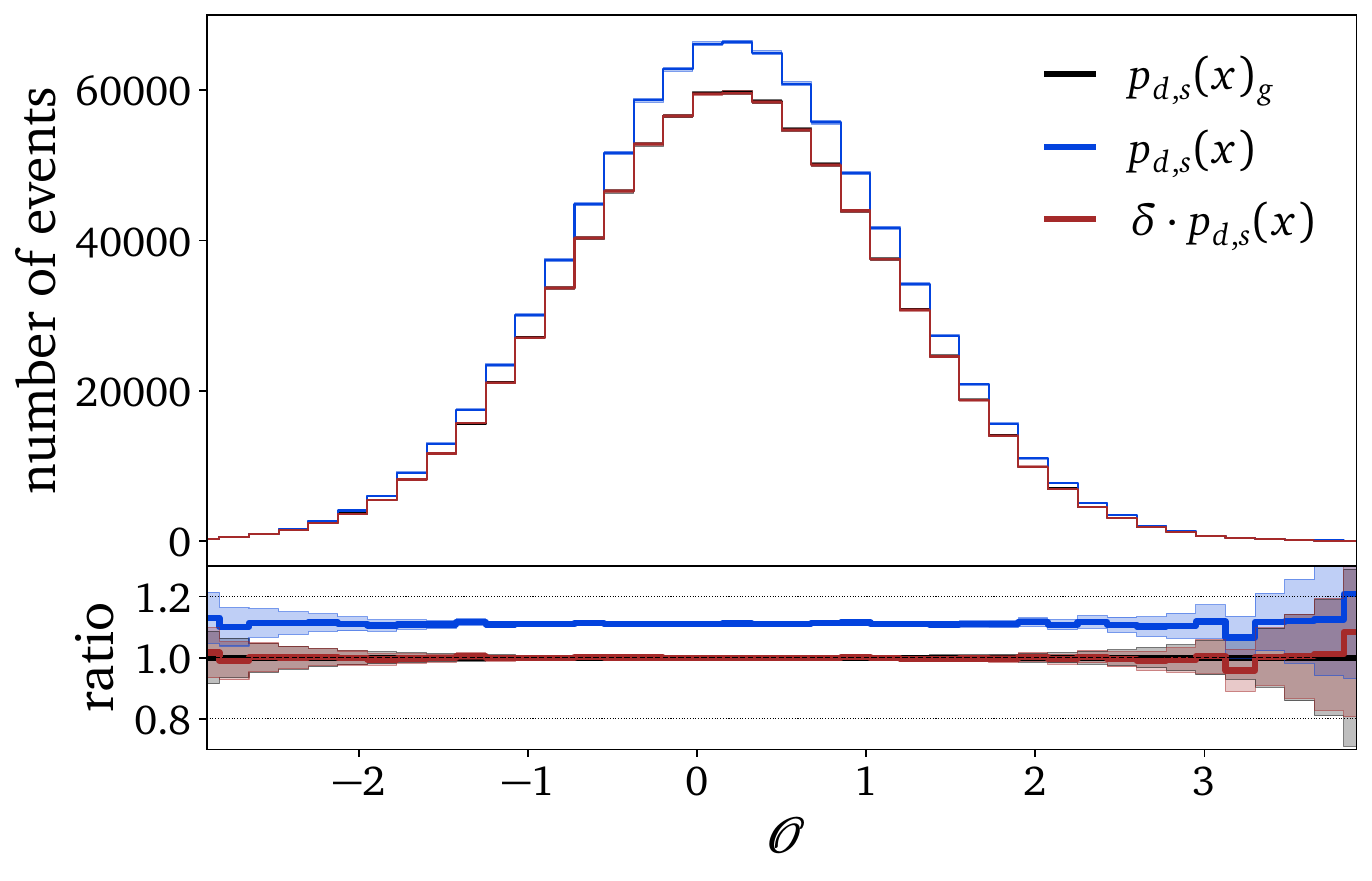} 
    \caption{Left: Comparison of truth reco-level signal data (black) with measured data = signal + background (blue) and the reweighted (red) or generated version (orange) of it, which should be data - background = signal. All events shown passed reco-level cuts. Right: Comparison of signal data passing reco-level selection (blue), truth reco-level signal data that also passed gen-level cuts (black), and the reweighted distribution accounting for gen-level cuts by the learned acceptance $\delta (x)$.}    
    \label{fig:toy_model_background}
\end{figure}
\begin{figure}[t]
    \centering
    \includegraphics[width=0.45\textwidth]{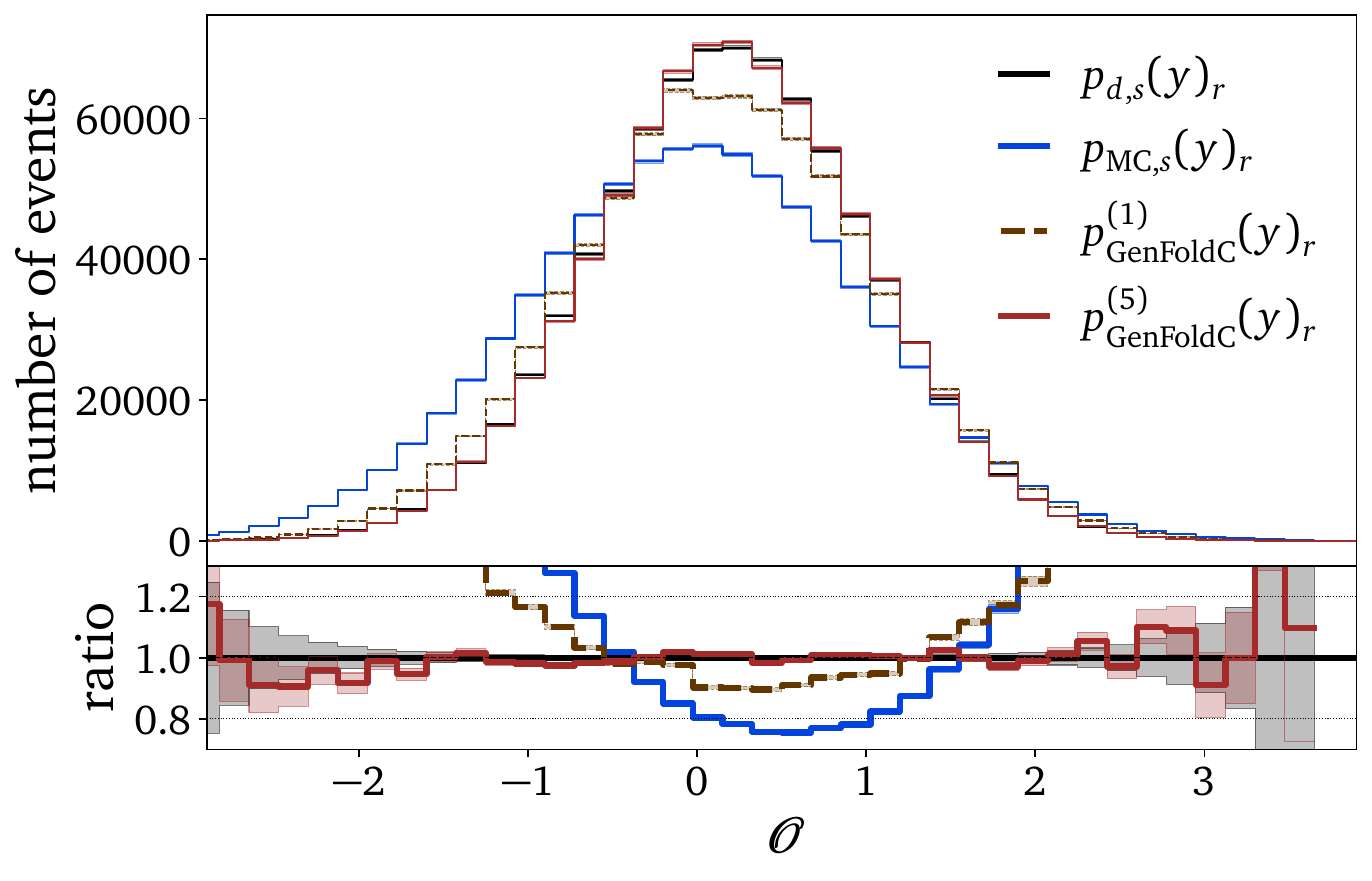}
    \hspace{0.5cm}
    \includegraphics[width=0.45\textwidth]{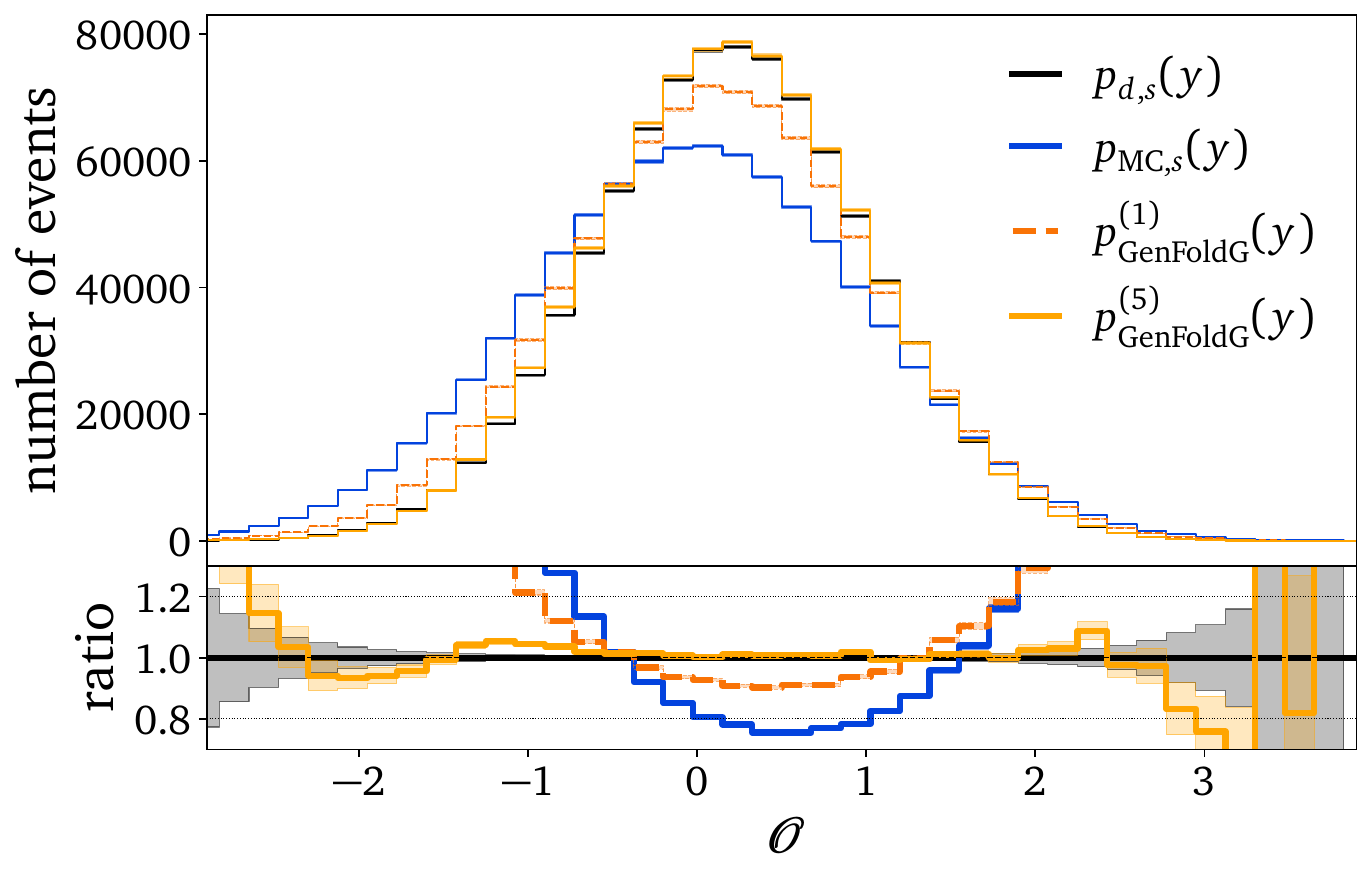}
    \caption{Results of the iterative generative unfolding after one iteration (dotted) and after five iterations (solid). We compare the unfolding against the gen-level distribution of the simulation (blue), and against the gen-level data distribution that passed gen- and reco-cuts (black). $p_{\text{GenFoldC}}(y)_{r}$ (left) and $p_{\text{GenFoldG}}(y)$ (right) is the result of our iterative unfolding. The former is provided implicitly with background weights and acceptance corrections, whereas the latter also includes efficiency corrections.}
    \label{fig:toy_model_final_prior}
\end{figure}
\begin{figure}[b]
    \centering    
    \includegraphics[width=0.45\textwidth]{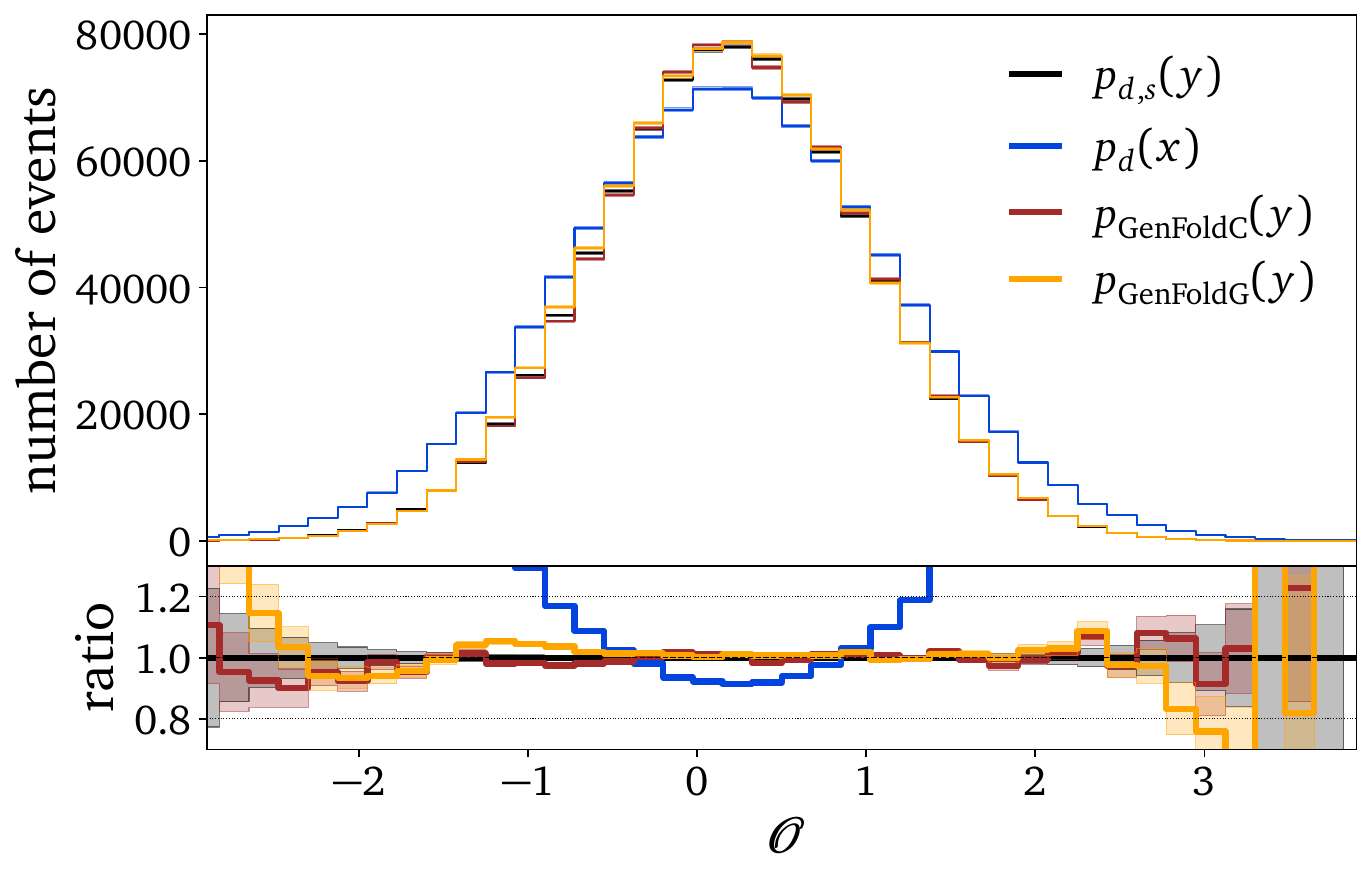}
    \caption{Distribution of the Gaussian toy example. Shown are the final generative unfolding of GenFoldC (red) and GenFoldG (orange), the truth gen-level signal data that passed gen-level selection (black) and the reco-level measurement containing background defined on the reco-level fiducial region (blue).}    
    \label{fig:toy_model_final_result}
\end{figure}
In this paper, all unfolding results are implemented with a generative conditional flow matching (CFM) network~\cite{lipman2023flowmatchinggenerativemodeling} as they have proven to be excellent tools for LHC related tasks~\cite{Butter:2023fov, Buhmann:2023zgc, Brehmer:2024yqw,Dreyer:2024bhs,Favaro:2024rle,Bothmann:2025lwg}. As for the remaining corrections, we chose to implement them either fully generator-based (GenFoldG), with additional CFMs, or fully classifier-based (GenFoldC). However, this is arbitrary, and the choice of generative or classifier-based correction 
can be made independently for each individual step. Details of the implementation of each neural network can be found in Tab.~\ref{tab:toy_c_hyperparameters} and Tab.~\ref{tab:toy_g_hyperparameters}.
We start by subtracting background and correcting for acceptance effects in Fig.~\ref{fig:toy_model_background}. We observe that the proposed algorithms in Sec.~\ref{sec:background_sub} and Sec.~\ref{sec:acceptance} are able to correct for these effects, respectively.
%
%
%
%
We proceed to train a generative network on paired MC samples, and evaluate the mapping on reco-level data. In Fig.~\ref{fig:toy_model_final_prior} we observe that the unfolding of the first iteration has a clear bias towards the reference MC simulation. To iteratively correct this bias, we chose either the suggested classifier- or generator-based algorithms. We notice a clear improvement between the first and the fifth iteration. Whereas after one iteration the unfolded results are off by more than 25\%, after five iterations the unfolded distribution matches the true $p_{d,s}(y)$ distribution to the percent level. 
As described in the previous section, the generator-base, iterative corrections provide a sample of unweighted unfolded events living in the fiducial gen-level phase space. For the classifier-based iteration style, we still need to correct for efficiency effects. This is done, by the efficiency classifier.
Lastly, we arrive at the final unfolding results in Fig.~\ref{fig:toy_model_final_result}. Both algorithms GenFoldC and GenFoldG adequately perform each correction step listed in Fig.~\ref{fig:unfolding_pipeline} and precisely estimate the truth gen-level data distribution $\pdsgg$.
%

\section{Unfolding Jet Observables}
\label{sec:omnifold_data}
Next, we consider a particle physics example where we build on the \textsc{OmniFold} dataset~\cite{Andreassen:2019cjw}. This dataset consists of simulated $pp \rightarrow Z+\text{jets}$ events at $\sqrt{s} = 14$~TeV, one generated with \Pythia~8.243~\cite{Sjostrand:2014zea} (in our case, used as the Monte Carlo simulation) and one generated with \Herwig~7.1.5~\cite{Bahr:2008pv, Bellm:2016voq, Bellm:2017bvx} (in our case, used as a stand-in for data). The detector response is simulated with \Delphes~3.4.2~\cite{deFavereau:2013fsa} using the CMS card. Particles are reconstructed with \Delphes, and jets of radius $R=0.4$ are clustered with \FastJet~3.3.2~\cite{Cacciari:2011ma} using the anti-$k_T$ algorithm~\cite{Cacciari:2008gp}. 
\begin{figure}[t]
    \centering    
    \includegraphics[width=0.45\textwidth, page=1]{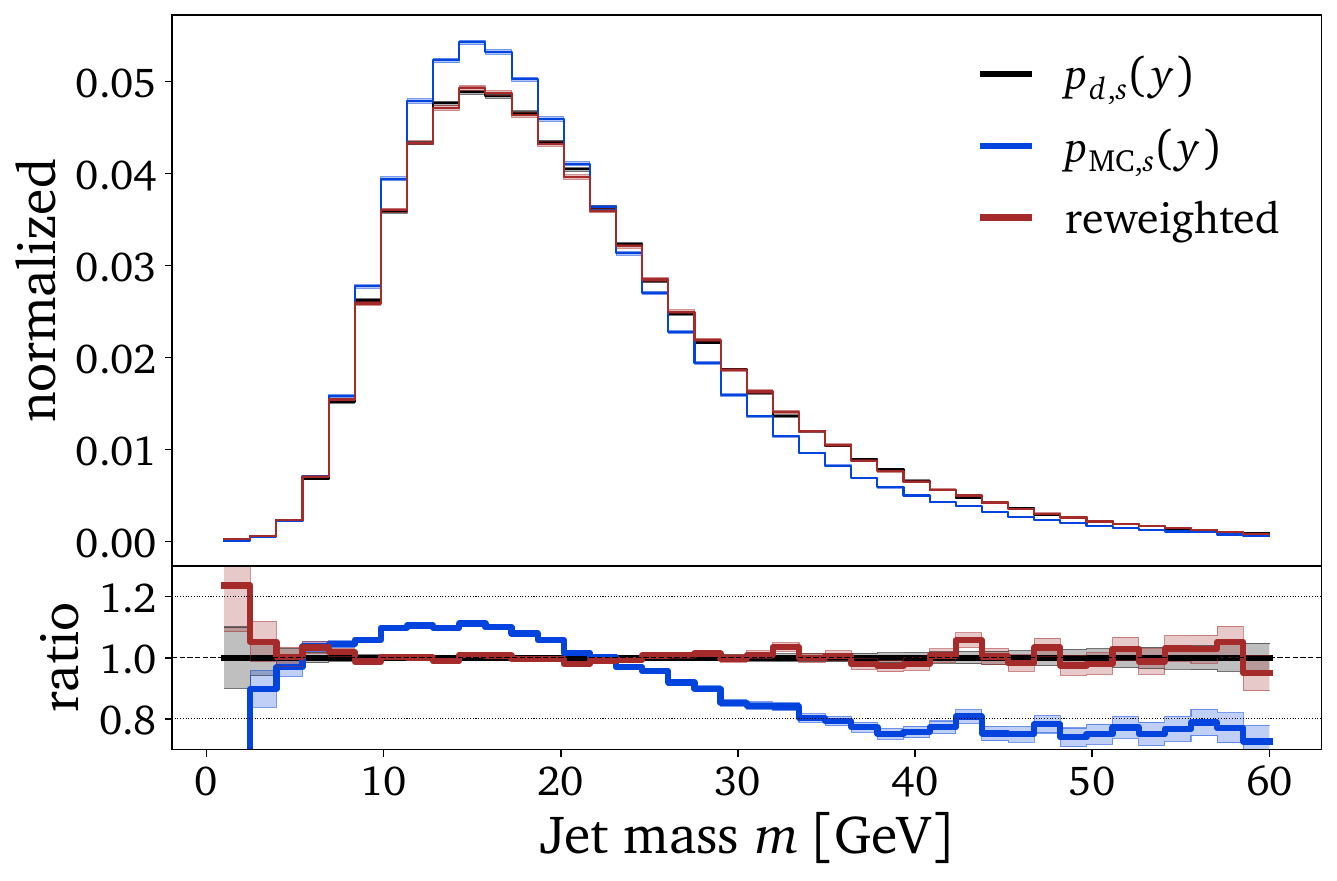}
    \hspace{0.5cm}
    \includegraphics[width=0.45\textwidth, page=2]{fig/omnifold/reweighted_pythia.pdf}
    \caption{Jet mass distribution, when reweighting gen-level MC (blue) to match the data distribution (black) once on gen-level (left) and once when weights are pulled to reco-level (right).}    
    \label{fig:reweighted_pythia}
\end{figure}
The detector response is expected to be universal at the level of detector inputs. However, when higher-level observables are considered, the detector response is no longer independent from the initial particle-level simulation. In practice, this would lead to a systematic uncertainty (see e.g. the hidden variable uncertainty in Ref.~\cite{ATLAS:2024xxl}), but we choose to remove it in order to more easily facilitate comparisons between methods. This is accomplished by first determining a reweighting function using a classifier. This reweighting function takes as inputs the same generator-level observables used in the unfolding task, learning to map simulation generated by the \Pythia simulation to the \Herwig simulation. After reweighting, all distributions at generator-level agree between distributions. When these weights are then also used to compare reconstruction-level observables, we see differences in the distributions, further evidence that even with the same detector simulation, the detector response is not universal. These results are shown in Fig.~\ref{fig:reweighted_pythia}. Additional details are provided in App.~\ref{app:different_forward_mappings}. 

%
%
\begin{figure}[t]
    \centering    
    \includegraphics[width=0.45\textwidth, page=1]{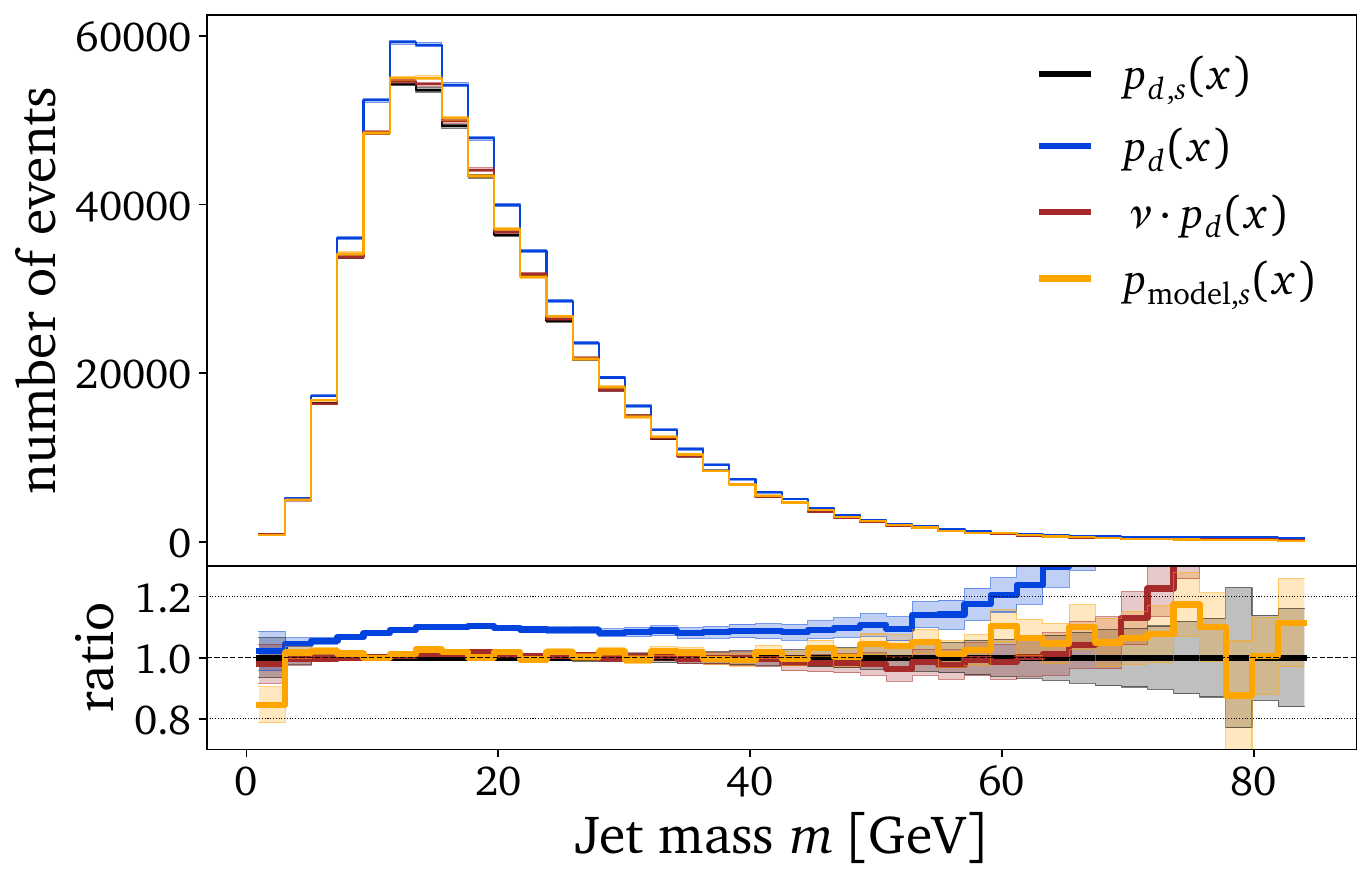} 
    \hspace{0.5cm}
    \includegraphics[width=0.45\textwidth, page=5]{fig/omnifold/background_subtraction.pdf}
    \caption{Subtracting $ZV$ background from $Z$+jets on reco-level in the jet mass and the N-subjetiness ratio $\tau_{12}$. The background accounts for around 8\% of the combined sample. We show the measured data (blue), signal data (black), classifier-based corrections (red) and generator-based subtraction (orange). All distributions are defined on the reco-level fiducial region.}
    \label{fig:bkg_omnifold}
\end{figure}
\begin{figure}[b]
    \centering    
    \includegraphics[width=0.45\textwidth, page=1]{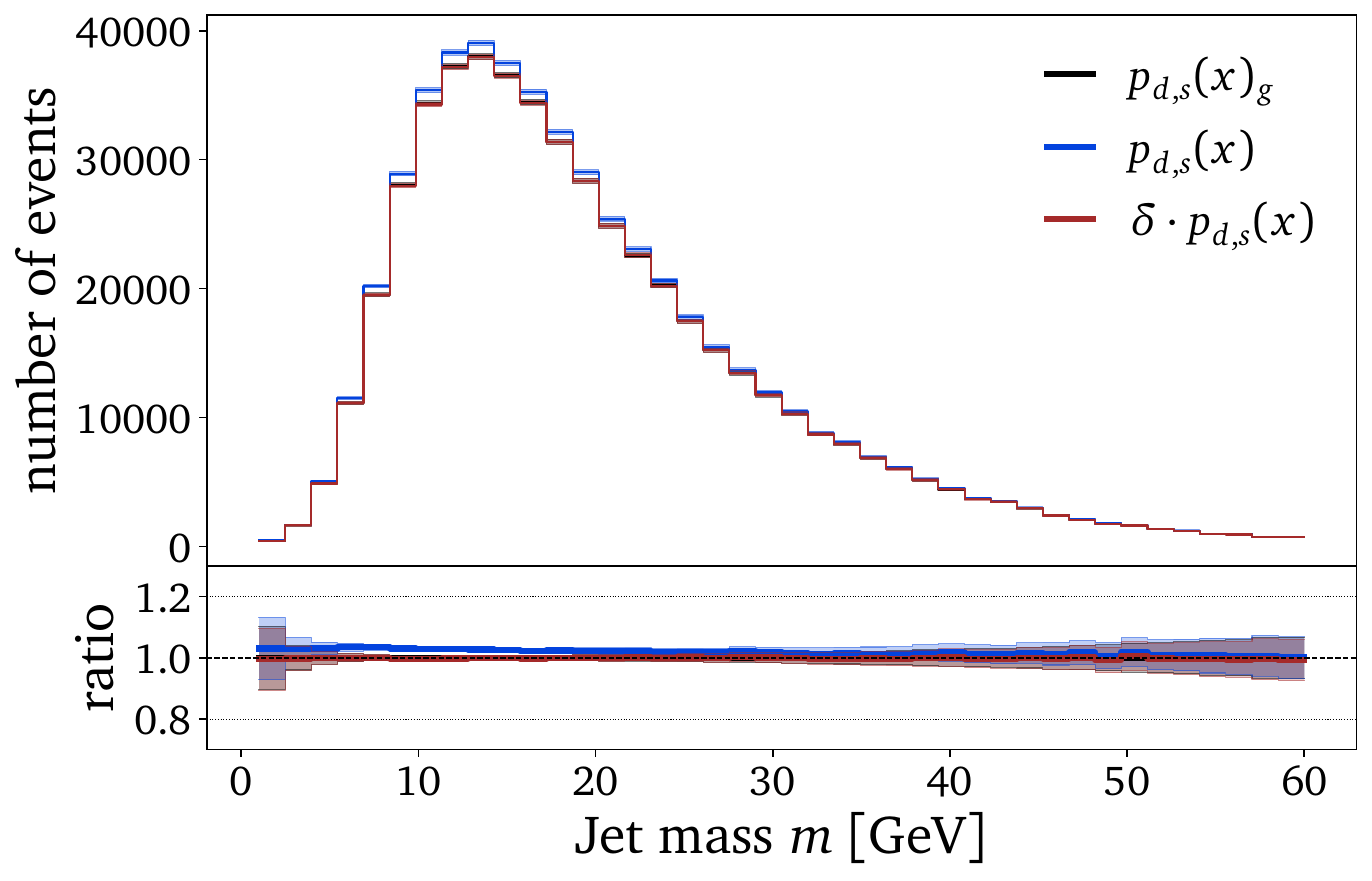} 
    \hspace{0.5cm}
    \includegraphics[width=0.45\textwidth, page=3]{fig/omnifold/acceptance.pdf} 
    \caption{Jet multiplicity $N$ and jet width $w$ at reco-level with gen-level cut (black), without gen-level cut (blue) and with learned correction (red).}
    \label{fig:acceptance_omnifold}
\end{figure}
\begin{figure}[t]
    \centering    
    \includegraphics[width=0.45\textwidth, page=1]{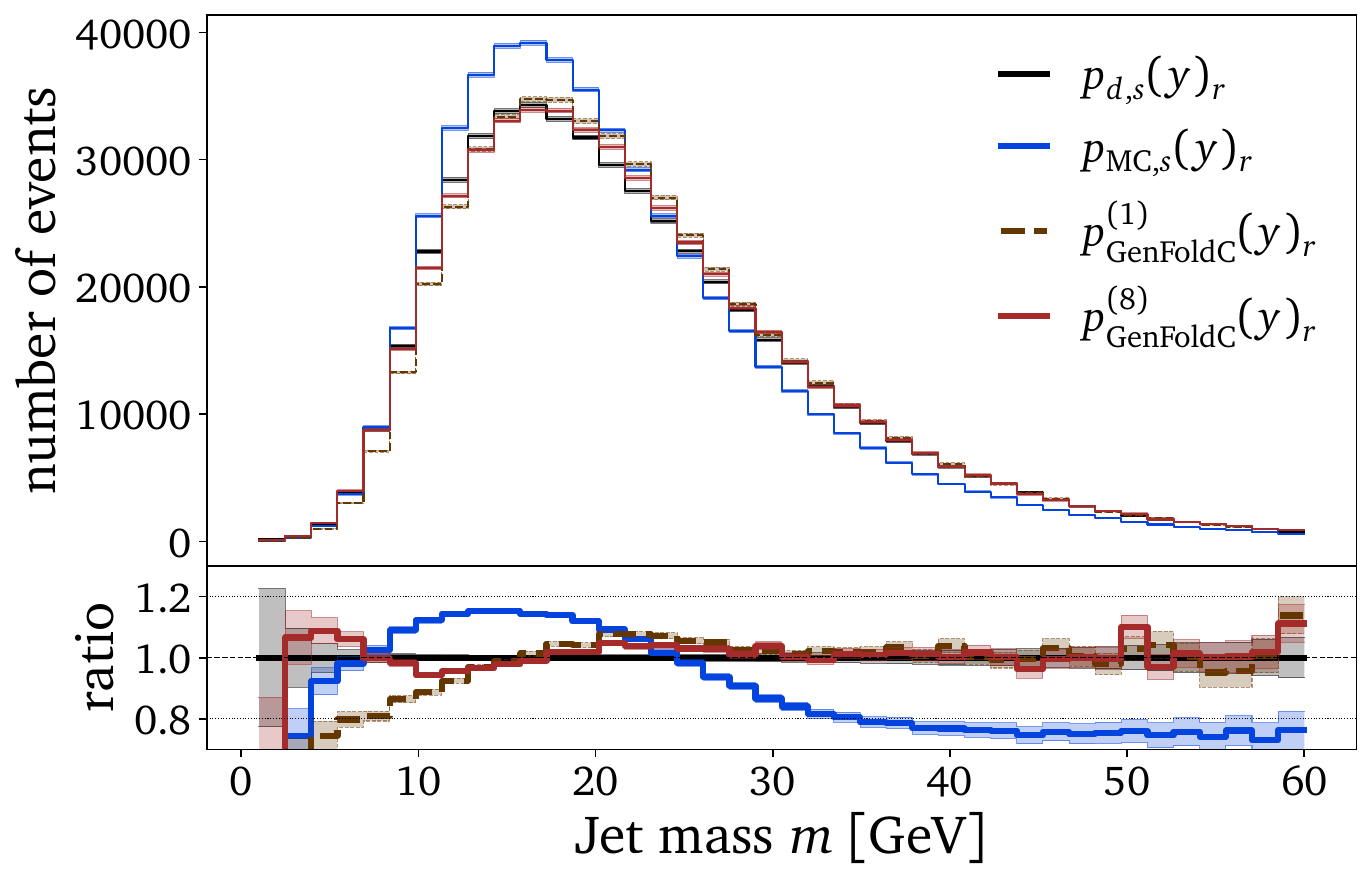} 
    \hspace{0.5cm}
    \includegraphics[width=0.45\textwidth, page=5]{fig/omnifold/genunf_c_iterations.pdf} \\
    \includegraphics[width=0.45\textwidth, page=1]{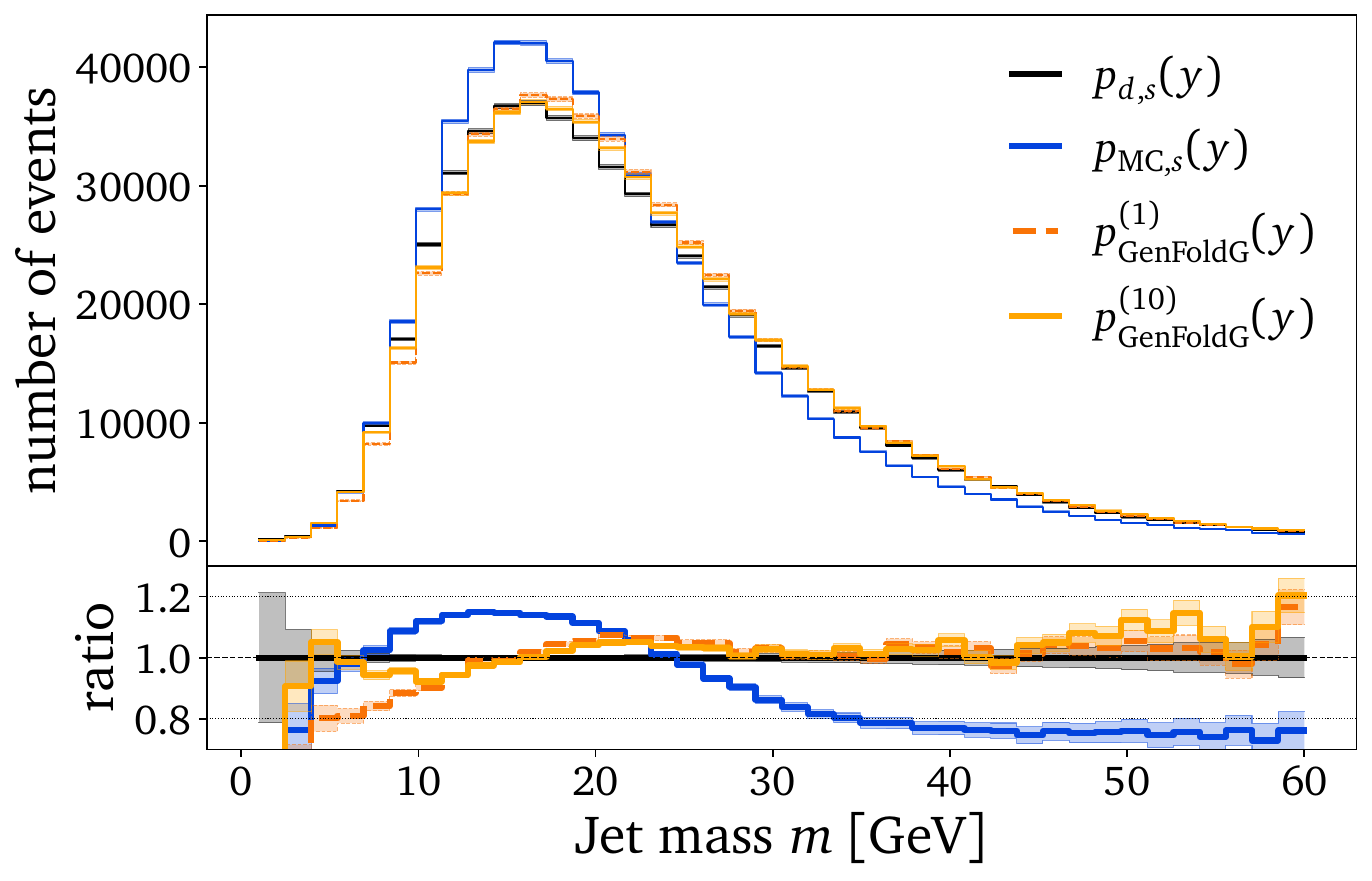} 
    \hspace{0.5cm}
    \includegraphics[width=0.45\textwidth, page=5]{fig/omnifold/genunf_g_iterations.pdf}
    \caption{Jet mass $m$ and N-subjettiness ratio $\tau_{21}$ distributions on gen-level. Shown are the MC distribution (blue), the truth data distribution (black) and the unfolded results after the first iteration (dotted) and at the end of the algorithm (solid). The upper row shows the results from GenFoldC, the lower row from GenFoldG.}
    \label{fig:prior_removal_omnifold}
\end{figure}
\begin{figure}[b]
    \centering    
    \includegraphics[width=0.45\textwidth, page=2]{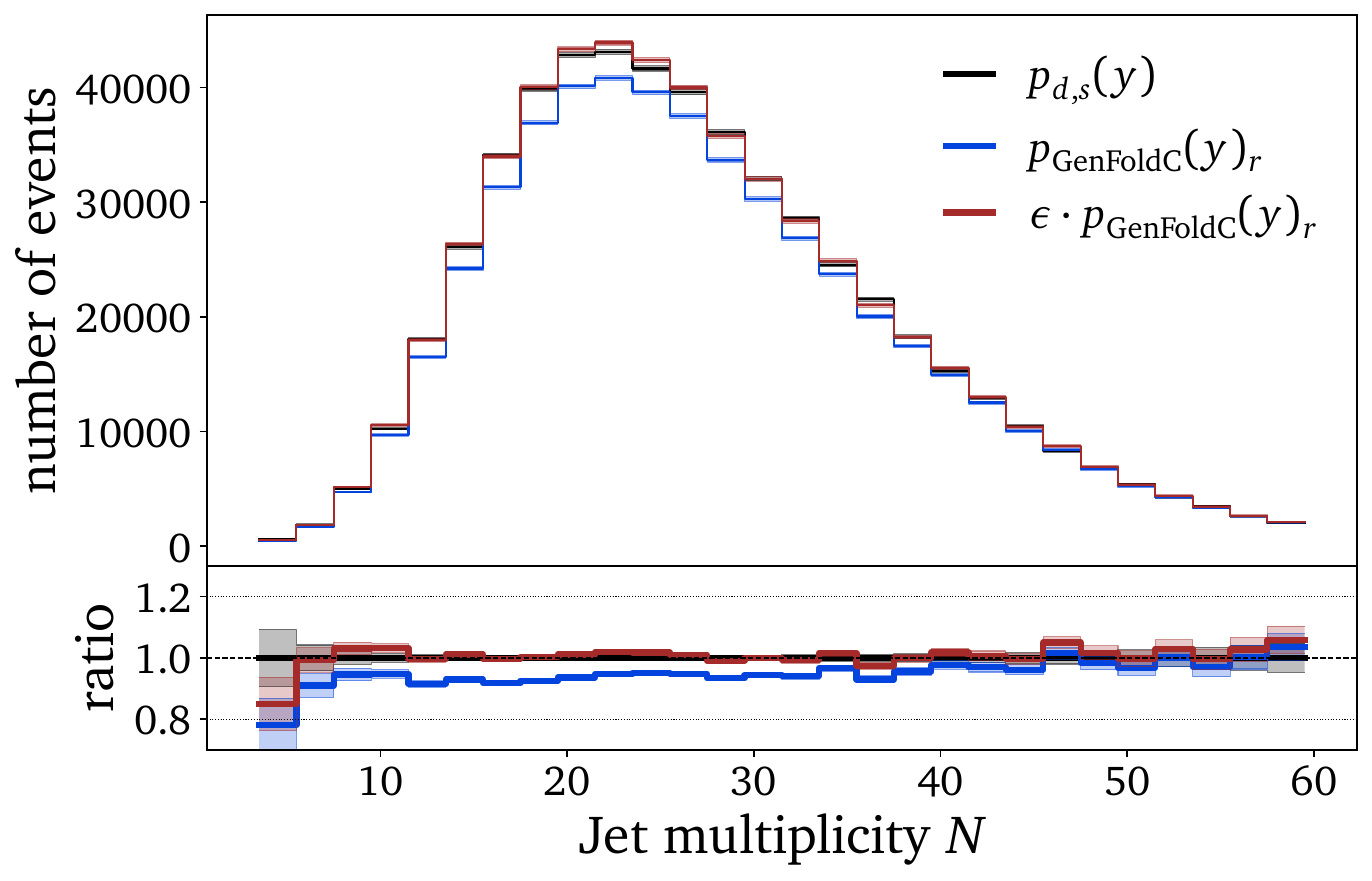} 
    \hspace{0.5cm}
    \includegraphics[width=0.45\textwidth, page=3]{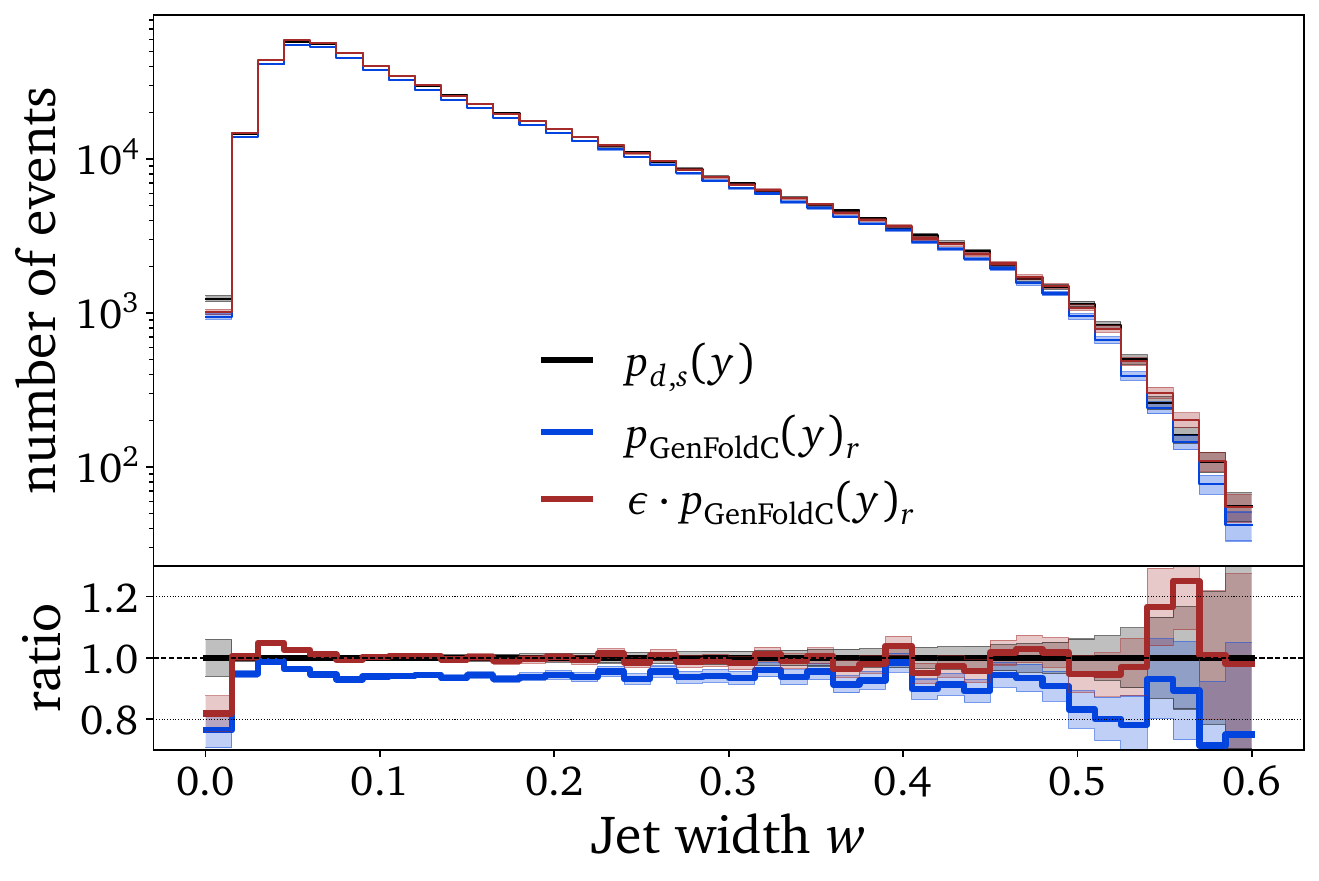} 
    \caption{Jet multiplicity and jet width distribution at gen-level (black), the unfolded distribution (blue) and the unfolded distribution when correcting for efficiency effects with a learned efficiency $\epsilon$ (red).}
    \label{fig:efficiency_omnifold}
\end{figure}
\begin{figure}[t]
    \centering    
    \includegraphics[width=0.45\textwidth, page=1]{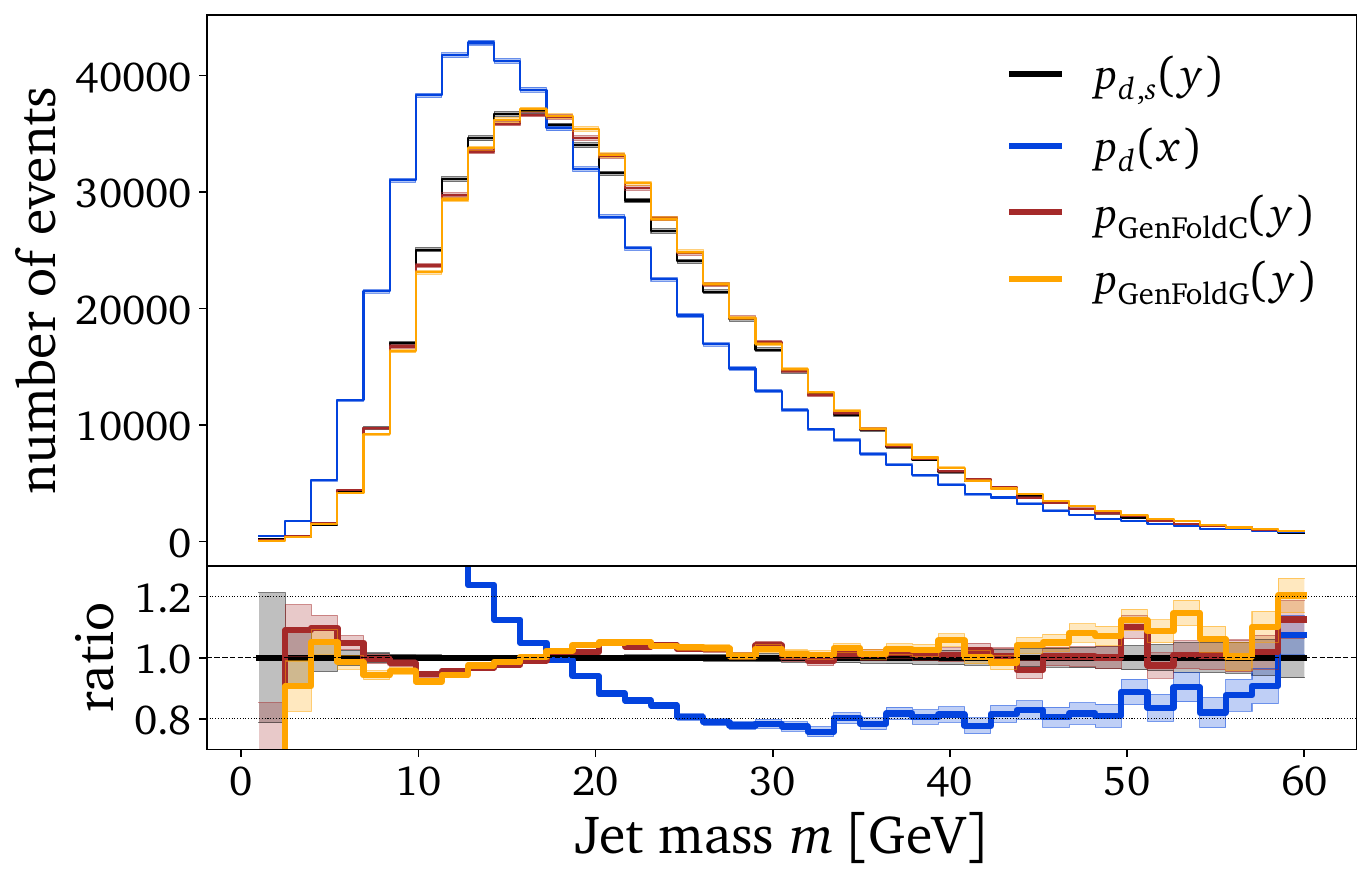} 
    \hspace{0.5cm}
    \includegraphics[width=0.45\textwidth, page=2]{fig/omnifold/final_unfolding.pdf} \\ 
    \includegraphics[width=0.45\textwidth, page=3]{fig/omnifold/final_unfolding.pdf} 
    \hspace{0.5cm}
    \includegraphics[width=0.45\textwidth, page=4]{fig/omnifold/final_unfolding.pdf} \\ 
    \includegraphics[width=0.45\textwidth, page=5]{fig/omnifold/final_unfolding.pdf} 
    \hspace{0.5cm}
    \includegraphics[width=0.45\textwidth, page=6]{fig/omnifold/final_unfolding.pdf}  
    \caption{Final 6-dimensional unfolding after background subtraction, acceptance corrections, iterative unfolding and efficiency correction for the GenFoldC (red) and GenFoldG (orange) algorithms. They are compared to the gen-level data distribution (black) and the reco-level measurement (blue).}
    \label{fig:final_omnifold}
\end{figure}
For the following results, we replace the \Herwig dataset with the reweighted \Pythia version to facilitate the comparison of the results.
Following the original studies using the \textsc{OmniFold} dataset we focus on the kinematic properties of the leading jet (in transverse momentum) of the event.
We chose the jet mass $m$, the jet multiplicity $N$, the jet width $w$, the log of the groomed mass $\log \rho$~\cite{Krohn:2009th,Ellis:2009me,Ellis:2009su,Dasgupta:2013ihk,Larkoski:2014wba}, the 2-subjetiness ratio $\tau_{12}$~\cite{Thaler:2010tr,Thaler:2011gf} and the groomed momentum fraction $z$~\cite{Dasgupta:2013ihk,Larkoski:2014wba} to be included during unfolding. 
Following Ref.~\cite{Milton:2025mug}, we study acceptance and efficiency effects by requiring the leading jet transverse momentum to be $p_T > 150 \;\text{GeV}$ on gen- and reco-level. 
We extend the dataset by including a background process for a measurement of strong $Z$ production.  For the background, we simulate $pp \rightarrow Z V \rightarrow \ell \ell qq$ events with \Pythia. The fragmentation products of both quarks can be clustered into one jet at these transverse momenta, leading to a similar signature as the signal process. Our signal-to-background ratio in data is approximately 8\%, slightly higher than the real value (see e.g. Ref.~\cite{ATLAS:2024xxl})
Details of the concrete implementations are listed in Tab.~\ref{tab:omnifold_c_hyperparameters} and Tab.~\ref{tab:omnifold_g_hyperparameters}.

Starting with the background subtraction in Fig.~\ref{fig:bkg_omnifold}, both classifier and generator-based subtraction algorithms are capable of correctly subtracting background in the high-statistics regions. 
There is some evidence for mis-modeling in the high-mass tail, where there are not so many events and the signal purity is relatively low.  It may be possible to mitigate such effects in the future by oversampling the background, by pre-training, or by using alternative statistical subtraction methods~\cite{Nachman:2025lid}.
Reaching a stable training setup for the background classifier has become increasingly harder. We attribute this to clustering effects of the background in the high-dimensional space. In such a scenario, the classifier is able to distinguish background from signal and the weighted BCE loss will become negative. 
%
%
%
%
In contrast to the Gaussian example, the kinematic selection that introduces the acceptance and efficiency effects in the physics example is strongly correlated between the gen- and reco-levels.  For example, even though on average 15\% fail the selection at gen-level, most of rejected events also do not pass the reco-level criteria. 
In total, only 2\% of the events pass reco-level but are not present at gen-level. Although small, we can still observe the effect of these corrections in Fig.~\ref{fig:acceptance_omnifold}.

We run both algorithms for eight and ten iterations, respectively, where we observe the convergence of the iterative procedure with minimal changes to the unfolded distributions for more iterations. To reduce the computational expense, later iterations are trained with a reduced number of epochs. In Fig.~\ref{fig:prior_removal_omnifold}, we compare the results obtained after a single unfolding iteration, where the choice of initial simulation has a strong impact on the unfolded results, and after convergence. Similarly to the Gaussian example, both methods obtain a clear improvement from the iterative procedure. We confirmed that at each iteration we always observe closure of the generative unfolding when unfolding the reference simulation. 
%
%
As for efficiency, within the \textsc{OmniFold} dataset 6\% of the events pass the gen-level selection but fail the reco-level criteria.
The efficiency classifier is able to correctly reweight the unfolded distribution $p_\text{GenUnfC}(y)_{r}$, such that it matches the expected gen-level data distribution $\pdsgg$ shown in Fig.~\ref{fig:efficiency_omnifold}.

After subtracting the background, applying acceptance and efficiency corrections, and running the iterative procedure to reduce the dependence on the simulation prior, we obtain the final results displayed in Fig.~\ref{fig:final_omnifold}. 
We obtain percent-level precision in most observables. Remaining differences between the unfolded and truth distributions are observed at extreme values of the N-subjetiness ratio and groomed mass. These disagreements are consistent with residual prior dependence that are not completely removed.  As discussed earlier, in practice, this would become a systematic uncertainty on the method.  Such an uncertainty results from the fundamental bias-variance trade off.  Future research may reduce this uncertainty for a given method or by ensembling a variety of methods with a similar precision.

\section{Conclusion and Outlook}
\label{sec:conc}

Given the rapid pace of theoretical and experimental developments in particle physics, it is increasingly essential to have data flexibility.  ML-based unfolding delivers this flexibility by allowing for observables to be chosen after the measurement instead of before, as with classical binned methods.  Additionally, seamlessly adding more dimensions has the potential to enhance precision.  However, given the ill-posed nature of unfolding, it is critical to have multiple methods with comparable sensitivity. 

In this paper, we have made iterative, generative unfolding methods analysis ready.  While previous works in this space focused on statistically removing resolution effects, we now also include procedures for dealing with background, acceptance, and efficiency effects.  Along the way, we have also introduced a new way to mitigate the dependence on the starting simulation.  We demonstrated our GenFoldC and GenFoldG algorithms on simulated Gaussian and collider physics examples.  In both cases, the methods are able to accurately account for all effects.

A number of challenges remain for research in unbinned unfolding.  While accurate, there are some places in the measured phase space where the methods introduce biases.  Some of these biases might be reducible with methodological innovation while some are fundamental to the bias-variance tradeoff.  Combining methods with similar sensitivity will be an interesting avenue to pursue in the future, both for improving and estimating precision.  
With the number of unbinned methods growing across particle physics, it will be exciting to see how new types of measurements will enhance our sensitivity to the Standard Model and what lies beyond.

\section*{Acknowledgments}

We thank Sascha Diefenbacher, Tilman Plehn, and Jesse Thaler for many useful conversations related to this paper.

AB, NH and SPS are supported by the BMBF Junior Group Generative Precision Networks for Particle Physics
(DLR 01IS22079). VM, and BN are supported by the U.S. Department of Energy (DOE), Office of Science under contract DE-AC02-05CH11231 and DE-AC02-76SF00515.  This research used resources of the National Energy Research Scientific Computing Center, a DOE Office of Science User Facility supported by the Office of Science of the U.S. Department of Energy under Contract No. DE-AC02-05CH11231 using NERSC awards HEP-ERCAP0021099 and HEP-ERCAP0028249.
\appendix
\section{Prior-dependent acceptance effects}
\label{app:prior_dependet}
\begin{figure}[b]
    \centering    
    \includegraphics[width=0.45\textwidth]{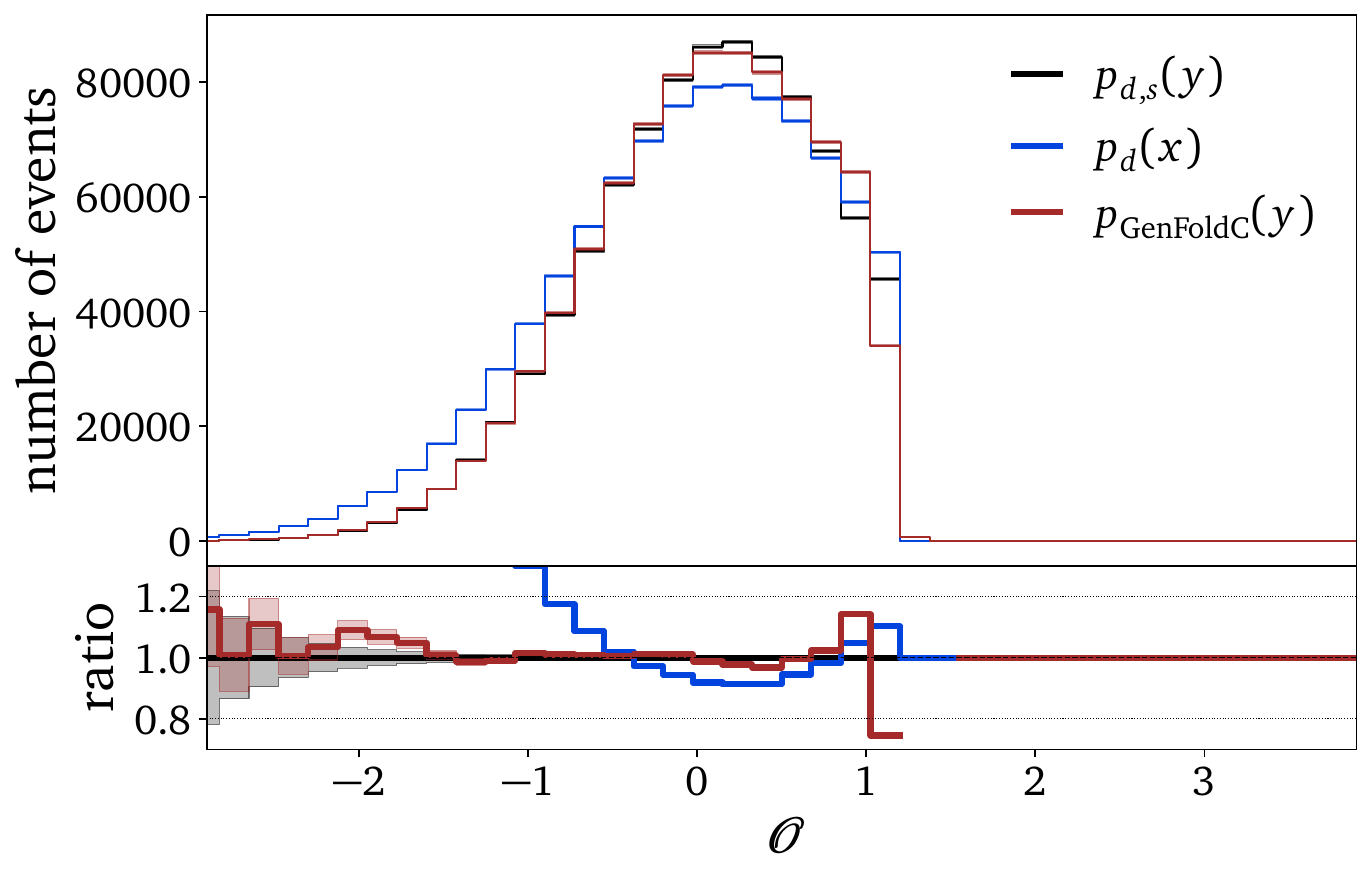}
    \hspace{0.5cm}
    \includegraphics[width=0.45\textwidth]{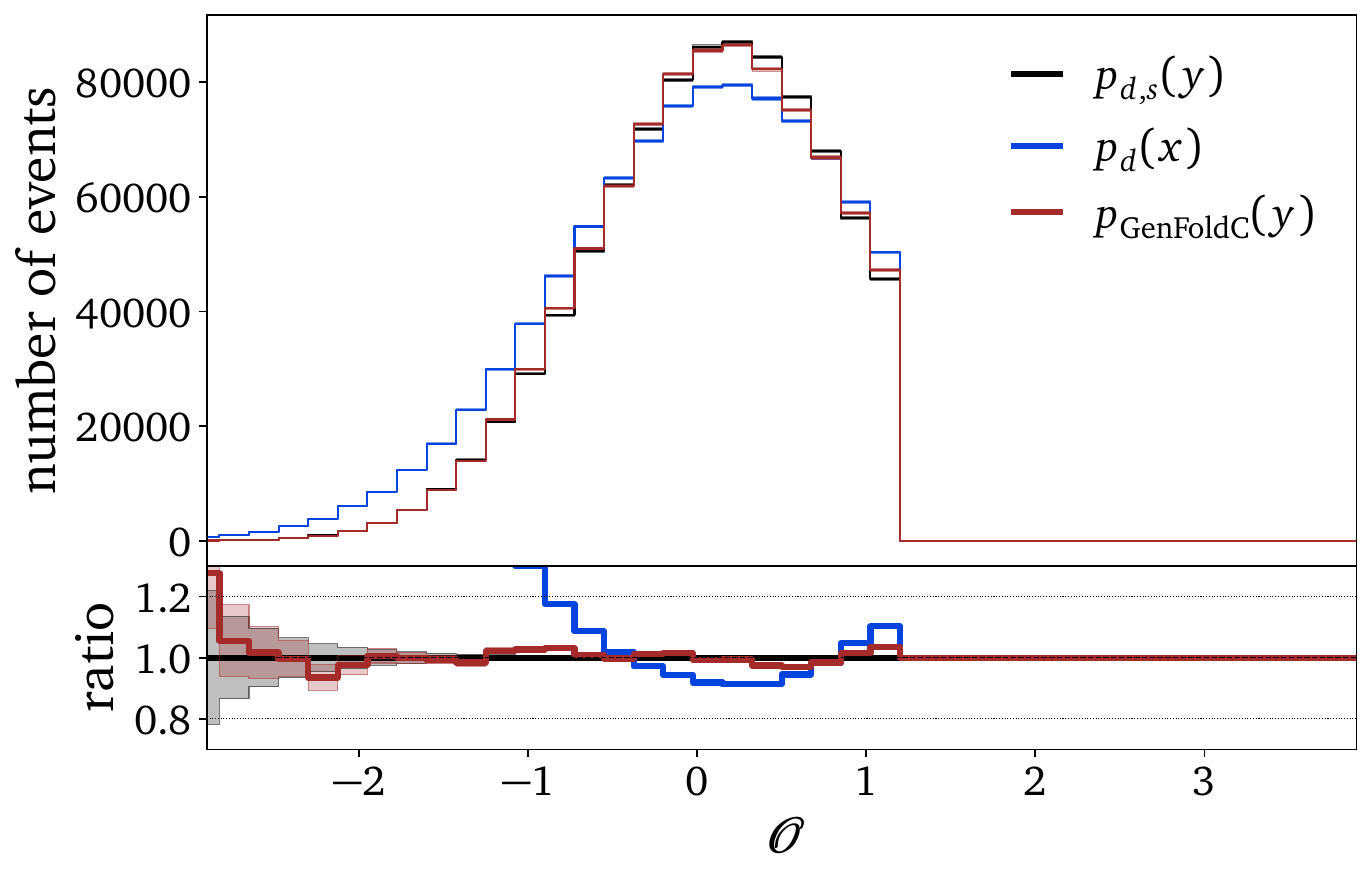}
    \caption{Final unfolded distribution (red), truth gen-level data (black) and reco-level measurement (blue) containing background, efficiency and acceptance effects of the Gaussian example introduced in Sec.~\ref{sec:gaussian_toy}. On the left, local acceptance effects are corrected for by Eq.~\eqref{eq:acceptance}, whereas on the right, local acceptance effects are corrected for by including sideband during the unfolding and a cut on the final unfolded distribution.}    
    \label{fig:prior_acceptance}
\end{figure}
Going back to Sec.~\ref{sec:acceptance} and looking more closely at $\delta(x)$, the acceptance ratio the classifier is trying to estimate is defined as 
\begin{align}
   \delta_{\text{true}}(x) = \int  dy\; p(y|x) \mathbbm{1}(y \in A) \; ,
\end{align}
where $A$ is the acceptance region in the gen-level phase space. The posterior distribution $p(y|x)$ introduces a prior dependence into the acceptance ratio. \\ 
In the Gaussian example and in the physics example, acceptance leads to the same statistical effects in simulation as in data. Therefore, this does not pose an issue.
To showcase an opposite scenario, we introduce a selection cut at $\mathcal{O} < 1.2$ on the one-dimensional Gaussian distribution. Here, we expect differences in acceptance effects between simulation and data because the selection cut is performed on the observable of interest and the unfolding problem is one-dimensional. Naively following the suggested classifier-based acceptance correction results in mismodelling the distribution close to the edge as visible in Fig.~\ref{fig:prior_acceptance}. 
In such scenarios, instead of correcting acceptance with a classifier, we propose to include sidebands and enforce the gen-level selection as a last step of the unfolding algorithm as done in Fig.~\ref{fig:prior_acceptance}. 
\section{Different forward mappings}
\label{app:different_forward_mappings}
As discussed in Sec.~\ref{sec:omnifold_data} and visible in Fig.~\ref{fig:reweighted_pythia}, reducing the \textsc{OmniFold}d dataset to a 6-dimensional summary statistics introduces differences in the forward mapping between simulations and (pseudo)-data. Hence, there will be a residual prior-dependence in the iterative unfolding, regardless whether we use generative unfolding or discriminative unfolding based on \textsc{OmniFold}.
\begin{figure}[t]
    \centering    
    \includegraphics[width=0.45\textwidth, page=1]{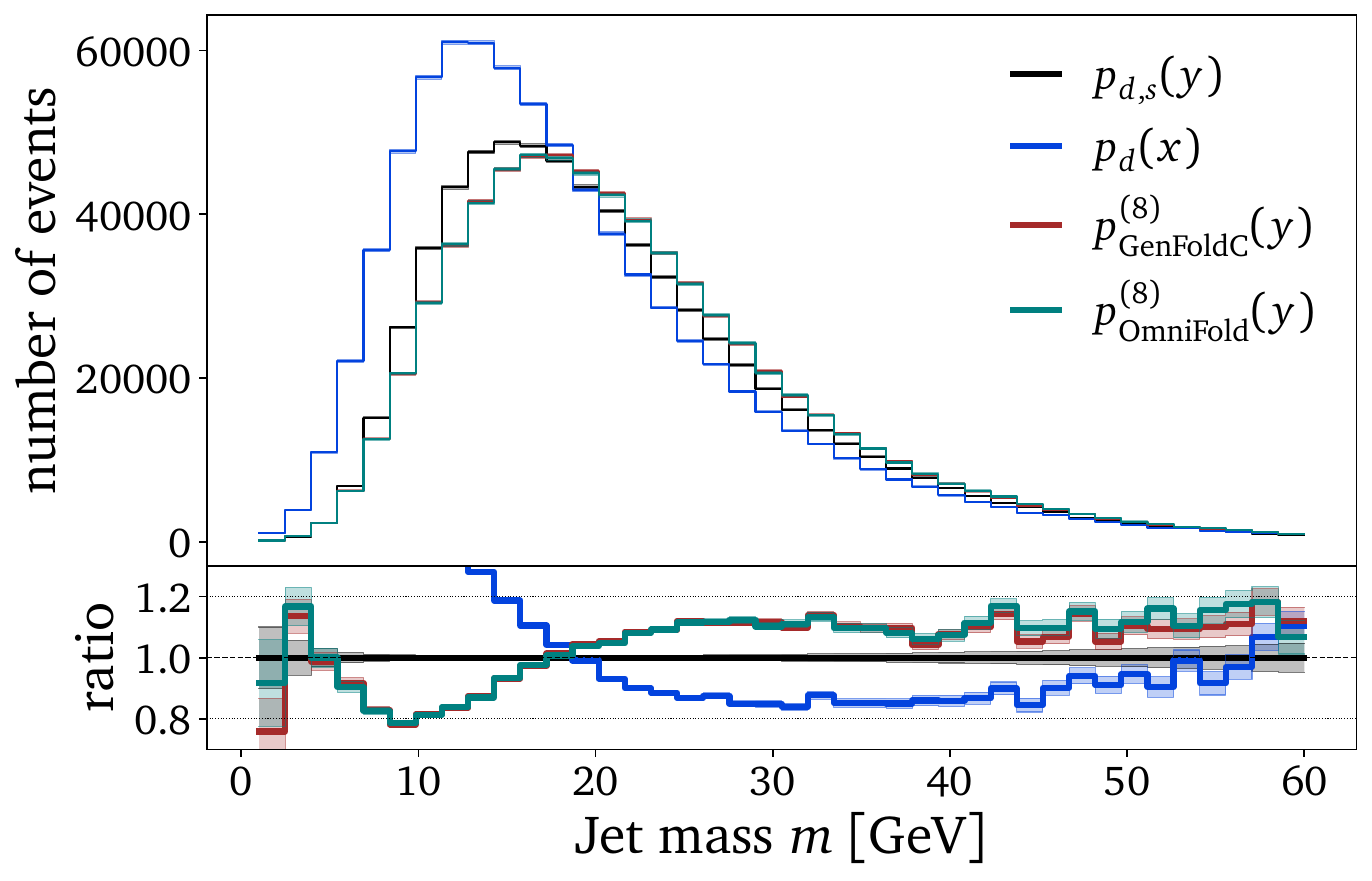}
    \hspace{0.5cm}
    \includegraphics[width=0.45\textwidth, page=2]{fig/omnifold/final_unfolding_og_omnifold.pdf}
    \caption{Truth gen-level data (\Herwig) in black, reco-level data (\Herwig) in blue and final unfolded jet mass distribution obtained with iterative generative unfolding in red and \textsc{OmniFold} in turquoise, without background, efficiency and acceptance effects after 8 iterations.}    \label{fig:og_omnifold}
\end{figure}
The deviations are most prominent in the jet mass as shown in Fig.~\ref{fig:og_omnifold}. The results are obtained following the same training procedure as in Sec.~\ref{sec:omnifold_data}, but without including background, efficiency and acceptance effects.
For the purpose of the presented study of this paper, we are not interested in investigating the prior removal for summary statistics further and leave this for future studies. 
As stated in the main body, we therefore use a reweighted \Pythia simulation rather than the original \Herwig version of the dataset. 
\section{Hyperparameters}
%
\begin{table}[!hb]
    \centering
    \begin{tabular}{l|cc} \toprule
        Parameter     &classifier & unfolder  \\ \midrule
        Epochs    & \multicolumn{2}{c}{20}    \\
        LR sched. & one-cycle & cosine \\
        LR    & \multicolumn{2}{c}{$10^{-4}$} \\
        Optimizer & \multicolumn{2}{c}{Adam}  \\
        Batch size & \multicolumn{2}{c}{128} \\
        \midrule
        Network  \\
        Intermediate dim   & \multicolumn{2}{c}{64}  \\
        Num layers   & \multicolumn{2}{c}{4} \\
        \bottomrule
    \end{tabular} 
    \caption{Hyperparameters to fully unfold Gaussian example used in Sec.~\ref{sec:pipeline} for GenFoldC. All classifiers are trained with the same hyperparameter setup.}
    \label{tab:toy_c_hyperparameters}
\end{table}

\begin{table}[!ht]
    \centering
    \begin{tabular}{l|ccc} \toprule
        Parameter     &classifiers & generators   \\ \midrule
        Epochs    & 10  & 100 \\
        LR sched. & one-cycle & cosine &\\
        LR    & $3\times10^{-5}$ & $10^{-4}$ \\
        Optimizer & \multicolumn{2}{c}{Adam} & \\
        Batch size & \multicolumn{2}{c}{128} &\\
        \midrule
        Network  \\
        Intermediate dim   & \multicolumn{2}{c}{64} & \\
        Num layers   & 4 & 3\\
        \bottomrule
    \end{tabular} 
    \caption{Hyperparameters to fully unfold Gaussian example used in Sec.~\ref{sec:pipeline} for GenFoldG. All classifiers are trained with the same hyperparameter setup.}
    \label{tab:toy_g_hyperparameters}
\end{table}

%
\begin{table}[!ht]
    \centering
    \begin{tabular}{l|cccc|c} \toprule
        Parameter     &bkg classifier & acc. classifier & eff. classifier& iter. classifier & unfolder \\
        \midrule
        Epochs    & \multicolumn{2}{c}{20}  &\multicolumn{2}{c}{50} & 400 (40) \\
        LR sched. & one-cycle  &\multicolumn{2}{c}{one-cycle}&one-cycle& cosine \\
        LR    &$10^{-4}$ &\multicolumn{2}{c}{$10^{-4}$}&$5 \cdot 10^{-4}$&$10^{-3}$ \\
        Optimizer & Adam & \multicolumn{2}{c}{Adam} & Adam& Adam \\
        Batch size & 10000&\multicolumn{2}{c}{2048}& 2048 &16384\\
        \midrule
        Network &&&  \\
        Intermediate dim   &64& \multicolumn{2}{c}{64} &128&512 \\
        Num layers   &4 & \multicolumn{2}{c}{3} &5&8\\
        \bottomrule
    \end{tabular} 
    \caption{Hyperparameters to fully unfold the \textsc{OmniFold} dataset used in Sec.~\ref{sec:omnifold_data}.} using GenFoldC.
    \label{tab:omnifold_c_hyperparameters}
\end{table}
%
\begin{table}[!hb]
    \centering
    \begin{tabular}{l|ccc} \toprule
        Parameter     &classifiers & generators   \\ \midrule
        Epochs    & 50  & 100 \\
        LR sched. & one-cycle & cosine &\\
        LR    & \multicolumn{2}{c}{$3\times10^{-5}$} &  \\
        Optimizer & \multicolumn{2}{c}{Adam} & \\
        Batch size & \multicolumn{2}{c}{256} &\\
        \midrule
        Network  \\
        Intermediate dim   & 64 & 512 \\
        Num layers   & 3 & 4\\
        \bottomrule
    \end{tabular} 
   \caption{Hyperparameters to fully unfold the \textsc{OmniFold} dataset used in Sec.~\ref{sec:omnifold_data}.} using GenFoldG.
    \label{tab:omnifold_g_hyperparameters}
\end{table}

\clearpage
\bibliographystyle{JHEP}
\bibliography{bibs/ben, bibs/ben_old, bibs/refs, bibs/tilman}
\end{document}